\documentclass[
prd,
superscriptaddress,
tightenlines,
showpacs,
showkeys,
nofootinbib,
aps,
amsfonts,
amssymb,
]{revtex4-1}

\usepackage{color}
\usepackage{amsmath}
\usepackage{amsfonts}
\usepackage{amssymb} 
\usepackage{graphicx}
\usepackage{epic}
\usepackage{eepic}
\usepackage{epsfig}
\usepackage{latexsym}
\usepackage{float}
\usepackage{slashed}
\usepackage[caption=false]{subfig}
\usepackage{tabularx}
\usepackage{bm}

\usepackage{geometry}
\begin{document}

\title{Prospects of heavy neutrino searches at future lepton colliders}

\author{Shankha Banerjee}
\email{shankha@hri.res.in}
\affiliation{Regional Centre for Accelerator-based Particle Physics,
Harish-Chandra Research Institute,
Chhatnag Road, Jhusi, Allahabad  211019, India}
\author{P. S. Bhupal Dev}
\email{bhupal.dev@hep.manchester.ac.uk}
\affiliation{Consortium for Fundamental Physics, School of Physics and Astronomy, 
University of Manchester, Manchester M13 9PL, United Kingdom}

\author{Alejandro Ibarra}
\email{alejandro.ibarra@ph.tum.de}
\affiliation{Physik-Department T30d, Technische Univertit\"{a}t M\"{u}nchen, 
James-Franck-Stra\ss e, D-85748 Garching, Germany}

\author{Tanumoy Mandal}
\email{tanumoymandal@hri.res.in}
\affiliation{Regional Centre for Accelerator-based Particle Physics,
Harish-Chandra Research Institute,
Chhatnag Road, Jhusi, Allahabad  211019, India}

\author{Manimala Mitra}
\email{manimala.mitra@durham.ac.uk}
\affiliation{Institute for Particle Physics Phenomenology, 
Durham University, Durham DH1 3LE, United Kingdom}


\begin{abstract}

We discuss the future prospects of heavy neutrino searches at next generation lepton colliders. In particular, we focus on the planned electron-positron colliders, operating in two different beam modes, namely, $e^+e^-$ and $e^-e^-$. In the $e^+e^-$ beam mode, we consider various production and decay modes of the heavy neutrino ($N$), and find that the final state with $e+2j+\slashed{E}$, arising from the $e^+e^-\to N\nu$ production mode, is the most promising channel. However, since this mode is insensitive to the Majorana nature of the heavy neutrinos, we also study a new production channel $e^+e^-\to Ne^\pm W^\mp$, which leads to a same-sign dilepton plus four jet final state, thus directly probing the lepton number violation in $e^+e^-$ colliders. In the $e^-e^-$ beam mode, we study the prospects of the lepton number violating process of $e^-e^-\to W^-W^-$, mediated by a heavy Majorana neutrino. We use both cut-based and multivariate analysis techniques to make a realistic calculation of the relevant signal and background events, including detector effects for a generic linear collider detector. We find that with the cut-based analysis, the light-heavy neutrino mixing parameter $|V_{eN}|^2$ can be probed down to $\sim 10^{-4}$  at 95\% C.L. for the heavy neutrino mass up to $400$ GeV or so at $\sqrt s=500$ GeV with $100~\rm{fb}^{-1}$ of integrated luminosity. For smaller mixing values, we show that a multivariate analysis can improve the signal significance by up to an order of magnitude. These limits will be at least an order of magnitude better than the current best limits from electroweak precision data, as well as the projected limits from $\sqrt s=14$ TeV LHC.

\end{abstract}

\preprint{jblml}

\pacs{12.60.-i, 14.60.St, 12.38.Qk}
\keywords{Beyond Standard Model, Neutrino Physics, $e^+e^-$ Experiments}

\maketitle 

\section{Introduction}\label{sec:1}

The observation of neutrino oscillations in solar, atmospheric, reactor and accelerator neutrino data (for a review, see e.g.~\cite{Agashe:2014kda}) has unequivocally established that at least two of the three active neutrinos have a non-zero mass and that individual lepton flavor is violated. This provides a conclusive experimental evidence for the existence of some new physics beyond the Standard Model (SM), and therefore, a precise understanding of the neutrino mass mechanism is an important step in unraveling the nature of new physics. 

The minimal renormalizable extension of the SM to explain the observed smallness of neutrino masses is the {\em neutrino portal}, defined by the Lagrangian
\begin{align}
-{\cal L} \ = \ h_{\ell \alpha} \bar{L}_\ell \widetilde{\Phi} N_\alpha + \frac{1}{2}M_{N_{\alpha \beta}} \bar{N}_\alpha^C N_\beta + {\rm H.c.} \; ,
\label{lagY}
\end{align}
where $L_\ell = (\nu_\ell, ~ \ell)_L^{\sf T}$ (with $\ell=e,\mu,\tau$) and $\Phi$ are the $SU(2)_L$ lepton and Higgs doublets respectively,  $\widetilde \Phi=i\sigma_2\Phi^*$ ($\sigma_2$ being the second Pauli matrix) is the hypercharge conjugate of $\Phi$, $N^C_\alpha\equiv N_\alpha^{\sf T}C^{-1}$ ($C$ being the charge conjugation matrix) denotes the charge conjugate field and $N_\alpha$ are SM gauge-singlet neutral fermions, usually known as the sterile neutrinos, since they can talk to the SM sector only via their mixing with the active neutrinos. In Eq.~\eqref{lagY}, $h_{\ell \alpha}$ are the dimensionless complex Yukawa couplings which, after the electroweak symmetry breaking, yield a Dirac mass matrix $M_D=hv$, $v$ being the Higgs vacuum expectation value. In addition, Eq.~\eqref{lagY} allows a Majorana mass term $M_N$, since the sterile neutrinos carry no SM gauge charge. The existence of these new interactions in Eq.~\eqref{lagY} leads to the following neutrino mass matrix in the flavor basis $\{\nu^C_\ell,N_\alpha\}$:
\begin{align}
{\cal M}_\nu \ = \ \begin{pmatrix}
0   & M_D \\
M_D^{\sf T} & M_N
\end{pmatrix} \; .
\label{Mnubig}
\end{align}
The active-sterile neutrino mixing is parametrized by $\xi \sim M_DM_N^{-1}$. In our subsequent discussion, we often denote the elements of this matrix as $V_{\ell N}$ (see Section~\ref{sec:2} for details). For $\|\xi\| \equiv \sqrt{{\rm Tr}(\xi^\dag \xi)} \ll 1$, the light neutrino masses and mixing are given by the diagonalization of the effective mass matrix 
\begin{align}
M_\nu \ \simeq \ - M_D M_N^{-1} M_D^{\sf T} \; , 
\label{Mnu}
\end{align}
whereas the heavy neutrino masses are of order $M_N$. This is known as the type-I seesaw mechanism~\cite{Minkowski:1977sc, mohapatra:1979ia, Yanagida:1979as, seesaw:1979, Schechter:1980gr}. 

From the above discussion, it is evident that there are two key aspects of the
seesaw  mechanism  that can  be  probed  experimentally: the  Majorana
mass $M_N$  of  the sterile  neutrinos  and  their mixing  $V_{\ell N}$ with the active ones. 
The  Majorana nature  of both active and sterile neutrinos can  in principle be probed  
via the lepton number violating (LNV) low-energy process of  neutrinoless          double          beta          decay
($0\nu\beta\beta$). 
However,   an observation of $0\nu\beta\beta$   does   not
necessarily  probe the  active-sterile mixing $V_{\ell N}$,  since
these  mixing  effects may not always  give  the  dominant contribution  
 over purely   left-or right-handed contributions to   the
$0\nu\beta\beta$ process; for a review, see e.g.~\cite{Rodejohann:2011mu}.  
Alternatively, a non-negligible value of $V_{\ell N}$ 
can be indirectly inferred from  non-unitarity of  the  light
neutrino mixing matrix~\cite{Antusch:2006vwa, abada:2007ux} as well as in observables for lepton flavor violation (LFV)~\cite{Dinh:2012bp, Alonso:2012ji}, lepton non-universality and electroweak precision tests~\cite{delAguila:2008pw, Akhmedov:2013hec, Basso:2013jka, Antusch:2014woa, Antusch:2015mia}.  
However, these low-energy  observables by themselves do not  prove the Majorana
nature  of neutrinos since models with pseudo-Dirac neutrinos can also give rise to large non-unitarity and LFV effects~\cite{Malinsky:2009gw, malinsky:2009df, Dev:2009aw, Forero:2011pc, LalAwasthi:2011aa, Humbert:2015epa}. 

On the other hand, the direct collider searches for heavy neutrinos at the energy frontier can provide a {\it simultaneous} probe of both the aspects of the seesaw mechanism, i.e. the Majorana nature of the heavy neutrinos $N$ and  their mixing with the active neutrinos, provided the seesaw scale is accessible to the available center of mass energy; for a review, see e.g.~\cite{Deppisch:2015qwa}.\footnote{For some examples of low-scale seesaw models with relatively large active-sterile neutrino mixing, which lead to observable effects at colliders as discussed here, see~\cite{Pilaftsis:1991ug, Buchmuller:1991ce, Gluza:2002vs, Pilaftsis:2004xx, Kersten:2007vk,  Xing:2009in, Gavela:2009cd, He:2009ua,  Adhikari:2010yt, Ibarra:2010xw, Deppisch:2010fr, Ibarra:2011xn, Mitra:2011qr, 
Dev:2009aw, Dev:2012sg, Dev:2012bd, Dev:2013oxa, Boucenna:2014zba}.} In particular, electroweak-scale heavy neutrinos can be produced on-shell at colliders with a cross section depending on the strength of their mixing with the active neutrinos in the minimal setup, while their subsequent decay pattern is governed by whether they are Majorana or Dirac particles. 
At a hadron collider, the `smoking gun' collider signal for heavy {\em Majorana} neutrinos is the same-sign dilepton final state with two jets and no missing transverse energy: $pp\to W^* \to N\ell^\pm \to \ell^\pm \ell^\pm jj$~\cite{Keung:1983uu, Datta:1993nm, Han:2006ip, Bray:2007ru, delAguila:2007em, Atre:2009rg}. Using this channel, both CMS and ATLAS experiments at the LHC have set direct limits on the light-heavy neutrino mixing $|V_{\ell N}|^2$ (with $\ell=e,\mu$) for heavy neutrino masses between 100 and 500 GeV~\cite{Chatrchyan:2012fla, Khachatryan:2015gha, ATLAS:2012yoa, klinger:2014}. With the run-II phase of the LHC with more energy and  higher luminosity, and including the infrared enhancement effects due to $t$-channel photon-mediated processes~\cite{Dev:2013wba}, these limits could in principle be extended for $M_N$ up to 1 TeV or so. 

For electroweak-scale {\em Dirac} neutrinos, the same-sign dilepton signal is suppressed, and the `smoking gun' signal at the LHC is the trilepton final state: $pp\to W^*\to N\ell^\pm \to \ell^\pm \ell^\mp \ell^\pm + \slashed{E}_T$~\cite{delAguila:2008cj, delAguila:2008hw, delAguila:2009bb, Chen:2011hc, Das:2012ze, Das:2014jxa}. Analyzing the trilepton data from $\sqrt s=8$ TeV LHC~\cite{Chatrchyan:2014aea}, similar limits on $|V_{\ell N}|^2$ (with $\ell=e,\mu$) have been derived for $M_N$ up to 500 GeV~\cite{Das:2014jxa}.

In light of vigorous discussions (see e.g.~\cite{Brock:2014tja}) on the possibility of a future lepton collider, such as ILC~\cite{Behnke:2013xla}, FCC-ee~\cite{Gomez-Ceballos:2013zzn}, CLIC~\cite{Aicheler:2012bya}, CEPC~\cite{Apyan:2015} and muon collider~\cite{Kaplan:2014xda}, it is worthwhile examining the prospects of heavy neutrino searches in the clean, almost background-free environment of a lepton collider. We recall that for heavy neutrino masses below the $Z$-boson threshold, using their possible production in the $Z$-boson decay $Z\to \nu_\ell N$ or $Z\to N\bar{\nu}_\ell$~\cite{Dittmar:1989yg}, and its subsequent neutral current (NC) or charged current (CC) decays, 95\% C.L. limits on $|V_{\ell N}|^2\lesssim 10^{-5}$ were obtained by L3~\cite{Adriani:1992pq} and DELPHI~\cite{Abreu:1996pa} collaborations from a reanalysis of the LEP data. A future high-luminosity $Z$-factory, such as the FCC-ee, will significantly enhance the sensitivity by several orders of magnitude~\cite{Blondel:2014bra}. Proposed fixed-target experiments such as SHiP~\cite{Bonivento:2013jag} can also dramatically improve the limits on $|V_{\ell N}|^2$, though only in the low-mass range between 0.5 and 10 GeV or so.

For heavy neutrino masses above the $Z$-boson threshold, a direct search was performed in $e^+e^-$ annihilation at LEP~\cite{Acciarri:1999qj, Achard:2001qv}, where a single heavy neutrino can be produced via its mixing with active neutrinos: $e^+e^-\to N\nu$~\cite{delAguila:1987nn, Buchmuller:1991tu, Djouadi:1993pe, Azuelos:1993qu, Gluza:1995js, Gluza:1997ts}, and can subsequently decay via CC, NC or Yukawa interactions with the SM $W$, $Z$ or Higgs ($H$) boson respectively: $N\to \ell^- W^+ ~(\ell^+W^-),~\nu (\bar{\nu}) Z,~\nu (\bar{\nu}) H$. Concentrating on the decay channel $N\to eW$ with $W\to$ jets, which would lead to a single isolated electron plus hadronic jets and missing energy, the L3 collaboration put a 95\% C.L. upper limit on the mixing parameter $|V_{eN}|^2$ in a heavy neutrino mass range between 80 and 205 GeV~\cite{Achard:2001qv}. Similar limits were derived recently~\cite{Antusch:2015mia} using the ALEPH data on $e^+e^-\to W^-W^+\to \bar{\nu}\ell^-\ell^+\nu$~\cite{Heister:2004wr}. However, these searches were kinematically limited by the maximum center-of-mass energy $\sqrt s=208$ GeV at LEP2. It is expected that future lepton colliders can significantly improve the sensitivity and extend the limits to higher heavy neutrino masses for both Dirac and Majorana cases. This is explicitly demonstrated here for the most promising channel $e^+e^-\to N\nu \to \ell jj \slashed{E}$\footnote{For a lepton collider, since the initial four-momenta of the particles are known, we can measure the full missing energy $\slashed{E}$ and not just the transverse component, as in the case of a hadron collider.} by performing a Monte Carlo simulation at $\sqrt s=350$ and 500 GeV ILC, including realistic detector effects and background estimates. We have used both cut-based and multivariate analysis (MVA) techniques to derive the projected sensitivity limits for a generic ILC detector~\cite{Behnke:2013lya}. We find that a light-heavy mixing of $|V_{eN}|^2$ down to $\sim 10^{-4}$ can be probed at 95\% C.L. for the heavy neutrino mass up to $400$ GeV or so at $\sqrt s=500$ GeV ILC with $100~\rm{fb}^{-1}$ integrated luminosity. Although for concreteness, we have performed our simulations for the ILC benchmark parameters, this study is also relevant in the context of other future lepton and photon colliders~\cite{Barletta:2014nka}. 

Note that neither the existing LEP search channels $e^+e^-\to N\nu \to \ell jj \slashed{E}$ nor $e^+e^-\to W^-W^+$ have LNV final states, and hence, it is not evident from these channels whether the heavy neutrinos are Majorana particles, as predicted by the usual seesaw mechanism. In principle, the Majorana or Dirac nature of the heavy neutrinos could be determined to some extent from the $e^+e^-\to N\nu \to \ell jj \slashed{E}$ mode using the distribution of the polar angle between the produced neutrino and the incoming electron~\cite{delAguila:2005pf}. Here, we propose an alternative clean way to  determine the Majorana nature using a new LNV signal at $e^+e^-$ colliders within the minimal seesaw: 
\begin{align}
e^+e^-\to N e^\pm W^\mp \to \ell^\pm W^\mp e^\pm  W^\mp \to \ell^\pm e^\pm + 4j \; ,
\end{align}
with same-sign dilepton plus four hadronic jets. 
Noting that the SM background for this process is extremely small, we perform a detailed detector-level simulation for the signal sensitivity and  find that the detection of 10 signal events for a mixing parameter $|V_{eN}|=0.04$ would require an integrated luminosity of at least $700 ~{\rm fb}^{-1}$, which is within reach of next generation lepton colliders.  

For completeness, we also present the heavy neutrino production cross sections for various other subdominant processes, such as the Higgsstrahlung process $e^+e^-\to ZH \to ZN\nu_\ell (\bar{\nu}_\ell)$ and the dominantly photon-mediated processes $e^+e^-\to N\ell^\pm e^\mp \nu_e (\bar{\nu}_e)$. We could also have the processes $e^\pm \gamma \to N W^\pm$ and $e^\pm \gamma \to N\ell^\pm \nu_e (\bar{\nu}_e)$, with a real photon coming from the electron or positron beam. These processes are of direct relevance for an electron-photon collider~\cite{Bray:2005wv}. 

Finally, taking advantage of the fact that it is possible to switch the beam configuration of an $e^+e^-$ machine from $e^+e^-$ to an $e^-e^-$ mode~\cite{Adolphsen:2013kya}, one can also look for an LNV signal in the process $e^-e^-\to W^-W^-\to 4j$ mediated by a $t$-channel Majorana neutrino~\cite{Gluza:1995ix, Belanger:1995nh, Ananthanarayan:1995cn, Greub:1996ct, Rodejohann:2010jh}. We make a detailed detector-level analysis of this LNV signal and find that due to negligible SM background for this process, it offers an interesting alternative way to probe LNV in the electron sector, even though the cross section is suppressed by $|V_{eN}|^4$. 

The plan of the paper is as follows: in Section~\ref{sec:2}, we discuss various possible production modes of the heavy neutrino at a lepton collider. In Section~\ref{sec:3}, we do a detailed event analysis using the cut-based techniques for the signal and background involving the final states with $e+2j+\slashed{E}$ and $e^\pm e^\pm+4j$ at the $e^+e^-$ collider and the LNV $4j$ final state  at the $e^-e^-$ collider. In Section~\ref{sec:4}, we present an MVA to further enhance the signal sensitivity for the most promising channel with $e+2j+\slashed{E}$ final state. Our conclusions are given in Section~\ref{sec:conclu}. 

\section{Heavy Neutrino Production at Lepton Collider} \label{sec:2}
In the minimal seesaw scenario, where the heavy neutrinos do not carry any gauge charge, the only way for them to communicate with the SM sector is through their mixing with the active neutrinos, as governed by the seesaw Lagrangian~\eqref{lagY}. To parametrize the light-heavy neutrino mixing, we first diagonalize the full neutrino mass matrix given in Eq.~\eqref{Mnubig} by a unitary mixing matrix: 
\begin{align}
{\cal V}^{\sf T}{\cal M}_\nu {\cal V} \ = \  {\rm diag}(m_i\: , M_j) \; ,
\label{diag}
\end{align}
where $m_i$ (with $i=1,2,3$) and $M_j$ (with $j=4,5,...$) are respectively the light and heavy neutrino mass eigenvalues. The unitary matrix ${\cal V}$ has an exact representation in terms of 
a dimensionless matrix $\xi$ (which depends on $M_D$ and $M_N$):~\cite{Korner:1992zk, Dev:2012sg} 
\begin{align}
{\cal V} = \left(\begin{array}{cc}
({\bf 1}+\xi^*\xi^{\sf T})^{-1/2} & \xi^*({\bf 1}+\xi^{\sf T}\xi^*)^{-1/2} \\
-\xi^{\sf T}({\bf 1}+\xi^*\xi^{\sf T})^{-1/2} & ({\bf 1}+\xi^{\sf T}\xi^*)^{-1/2}
\end{array}  \right)
\left(\begin{array}{cc}
{\cal U} & {\bf 0} \\
{\bf 0} & {\cal U}'
\end{array}\right) \; ,
\label{V}
\end{align}
where ${\cal U}$, ${\cal U}'$ are the  unitary matrices diagonalizing the individual light and heavy neutrino mass matrices $M_\nu$ and $M_N$ respectively. 

Now using Eqs. (\ref{diag}) and (\ref{V}), the light neutrino flavor eigenstates $\nu_\ell$ can be related to the mass eigenstates $\widehat{\nu}_i$ and $\widehat{N}_j$ as follows: 
\begin{align}
\nu_\ell \ = \ \left[({\bf 1}+\xi^*\xi^{\sf T})^{-1/2}\right]_{\ell \ell'} {\cal U}_{\ell' i} \widehat{\nu}_i + \left[\xi^*({\bf 1}+\xi^{\sf T}\xi^*)^{-1/2}\right]_{\ell \ell''}{\cal U}'_{\ell'' j}\widehat{N}_j 
\ \equiv \ U_{\ell i} \widehat{\nu}_i + V_{\ell j} \widehat{N}_j\; ,
\end{align}
where the first term on the right-hand side (RHS) measures the non-unitarity of the Pontecorvo-Maki-Nakagawa-Sakata mixing matrix, 
and the second term determines the size of the light-heavy neutrino mixing in 
CC and NC interactions involving neutrinos.  In the charged-lepton mass diagonal basis, the CC interaction relevant for the production and decay of heavy neutrinos is given by 
\begin{align}
-\mathcal{L}_{CC} \  = \ \frac{g}{\sqrt{2}} W^-_{\mu}\bar{\ell} \gamma^{\mu} P_L \nu_\ell + {\rm H.c.} 
\  = \ \frac{g}{\sqrt{2}} W^-_{\mu}\bar{\ell} \gamma^{\mu} P_L  \left(U_{\ell i} \widehat{\nu_i}+ V_{\ell j} \widehat{N}_j \right) + {\rm H.c.} \; ,
\label{CC}
\end{align}
where $P_L = (1- \gamma^5)/2$ is the left-chirality projection operator. Similarly, the NC interaction is given by 
\begin{align}
-\mathcal{L}_{NC} \ = \  \frac{g}{2 \cos\theta_w}  Z_{\mu} \bar{\nu}_{\ell} \gamma^\mu P_L \nu_{\ell} 
 \ = \ & \frac{g}{2 \cos\theta_w}  Z_{\mu} \bigg[ 
(U^\dag U)_{ij}  \widehat{\bar{\nu}}_i \gamma^{\mu} P_L \widehat{\nu}_j 
 +  (V^\dag V)_{ij}\widehat{\bar{N}}_i \gamma^{\mu} P_L \widehat{N}_j  \nonumber \\ 
 & \qquad \qquad  
+ \left\{(U^\dag V)_{ij}\widehat{\bar{\nu}}_i \gamma^{\mu} P_L \widehat{N}_j 
  + {\rm H.c.} \right\} 
\bigg] , 
\label{NC}
\end{align}
 where $\theta_w$ is the weak mixing angle. For our subsequent discussion, we use a model-independent phenomenological approach, parametrized by a single heavy neutrino mass scale $M_N$, assuming that any other heavy neutrinos present in the system are sufficiently heavy and do not affect our analysis. Henceforth, we denote the light-heavy mixing parameter simply as 
$V_{\ell N}$, without explicitly writing the heavy neutrino flavor index. This enables us to derive generic collider exclusion/sensitivity limits in the $(M_N,V_{\ell N})$ parameter space, without referring to any particular low-scale seesaw model-building aspects. 

\begin{figure}[t]
\centering
\includegraphics[scale=0.75]{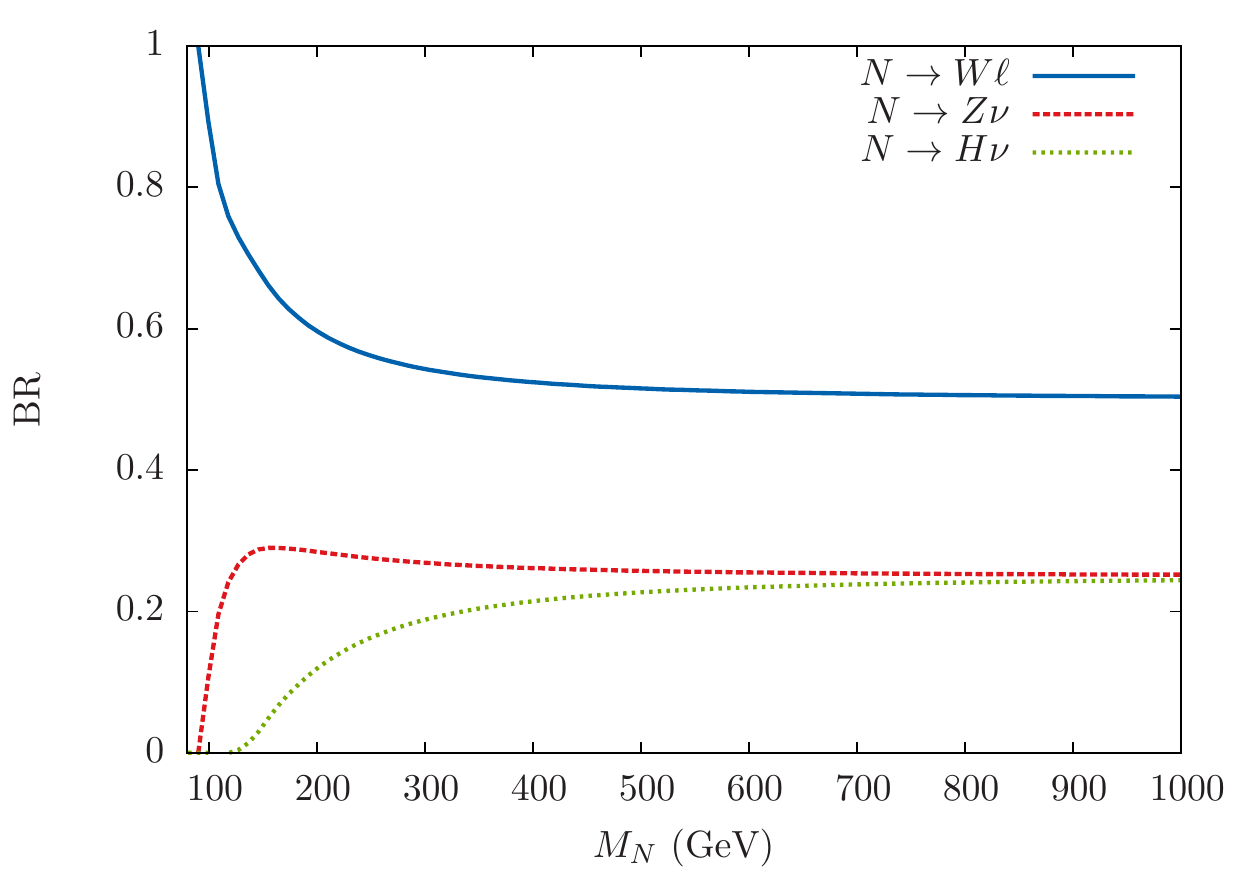}
\caption{The branching ratios (BR) of the heavy Majorana neutrino decay modes $N \to W^\pm \ell^\mp,~Z \nu(\bar{\nu}),~ H \nu (\bar{\nu})$ above the $Z$-boson threshold in the minimal seesaw model. }\label{fig:BR}
\end{figure}

From \eqref{CC} and \eqref{NC}, we see that the cross section of heavy neutrino production in $e^+e^-$ collision will be proportional to the mixing parameter $|V_{\ell N}|^n$, where $n=2$ or 4, depending on the process considered (see below). Similarly, 
for the two-body decays of the heavy neutrino to the SM final states $\ell^\pm W^\mp$, $Z\nu(\bar{\nu})$ and $H\nu(\bar{\nu})$  (if kinematically allowed), the corresponding partial decay widths are also proportional to $|V_{\ell N}|^2$:  
\begin{align}
\Gamma(N\to \ell^- W^+) & \ = \ \frac{g^2}{64\pi}|V_{\ell N}|^2\frac{M_N^3}{M_W^2}\left(1-\frac{M_W^2}{M_N^2}\right)^2\left(1+2\frac{M_W^2}{M_N^2}\right), \label{partial1}\\
\Gamma(N\to \nu_\ell Z) & \ = \ \frac{g^2}{128\pi}|V_{\ell N}|^2\frac{M_N^3}{M_W^2}\left(1-\frac{M_Z^2}{M_N^2}\right)^2
\left(1+2\frac{M_Z^2}{M_N^2}\right), \label{partial2} \\
\Gamma(N\to \nu_\ell H) & \ = \ \frac{g^2}{128\pi}|V_{\ell N}|^2\frac{M_N^3}{M_W^2}\left(1-\frac{M_H^2}{M_N^2}\right)^2. \label{partial3}
\end{align}
Note that the total decay width for a heavy Majorana neutrino will be twice the sum of the partial decay widths shown in Eqs.~\eqref{partial1}-\eqref{partial3}, after taking into account the charge conjugate processes. 
The corresponding branching ratios (BR) are shown in Figure~\ref{fig:BR}. For larger values of $M_N$, the branching ratios follow ${\rm BR}(\ell W):{\rm BR}(Z \nu):{\rm BR}(H \nu) \simeq 2:1:1$. Note that BR($N\to \ell^-W^+$) in the heavy Dirac neutrino case is twice as large compared to the heavy Majorana neutrino case, which has equal probability to decay into either $\ell^-W^+$ or $\ell^+W^-$. 

In what follows, we consider various heavy neutrino production channels at a lepton collider and numerically compute their production cross sections for given heavy neutrino mass and center of mass energy.\footnote{For general analytic expressions for the production cross section, see~\cite{Djouadi:1993pe}.} For our subsequent collider analysis, we have implemented the relevant Lagrangian terms as shown in Eqs. (\ref{CC}) and (\ref{NC}) for heavy Majorana neutrino interactions in~\textsc{FeynRules2.0} ~\cite{Christensen:2008py}, and generated the Universal FeynRules Output (UFO)~\cite{Degrande:2011ua} model files. These UFO model files have been used in the 
the \textsc{Monte-Carlo} (MC) event generator~\textsc{MadGraph5}~\cite{Alwall:2011uj} to generate the parton-level cross sections shown in this section, as well as the signal and background events for the analysis in Section~\ref{sec:3}. The following basic trigger cuts on lepton transverse momentum ($p_T^\ell$), pseudo-rapidity ($\eta^\ell$) and radial distance ($\Delta R^{\ell\ell}$) were used to identify the leptons in the final state: 
\begin{align}
p_T^\ell \ > \ 20~{\rm GeV}, \qquad |\eta^{\ell}| \ < \ 2.5,\qquad  \Delta R^{\ell \ell} \ > \ 0.4 \; .
\label{lepton_TC}
\end{align}

For our numerical simulation, we consider two benchmark values for the center of mass energy $\sqrt s=350$ and 500 GeV. Also, we focus on heavy neutrino masses only above the $Z$-threshold. For $M_N<M_Z$, the existing LEP limits from $Z$-decay are quite stringent~\cite{Adriani:1992pq, Abreu:1996pa} and one needs a high-luminosity $Z$-factory~\cite{Blondel:2014bra, Abada:2014cca} to significantly increase the sensitivity. 

\subsection{$\bm{e^+e^-\to N\nu_\ell}$}\label{sec:2.1}

\begin{figure}
\centering
\subfloat{
\begin{tabular}{cc}
\includegraphics[scale=0.55]{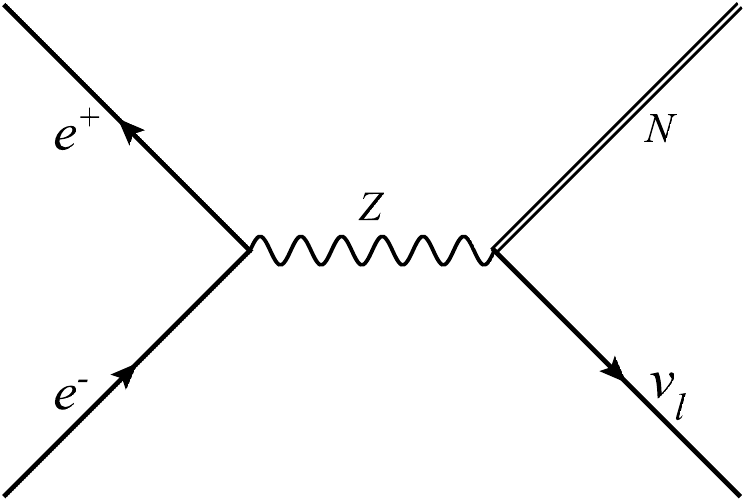} & \hspace{1cm}
\includegraphics[scale=0.55]{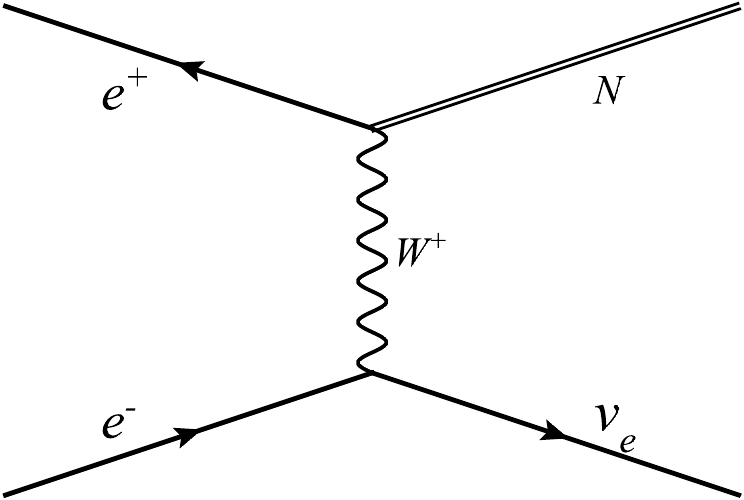} \\
(a) & \hspace{1cm} (b)
\end{tabular}}
\caption{\label{fig:FD}(a) The $s$-channel and (b) $t$-channel Feynman diagrams for the process $e^+e^-\to N\nu_\ell$.}
\end{figure}

\begin{figure}
\centering
\subfloat{
\begin{tabular}{cc}
\includegraphics[scale=0.6]{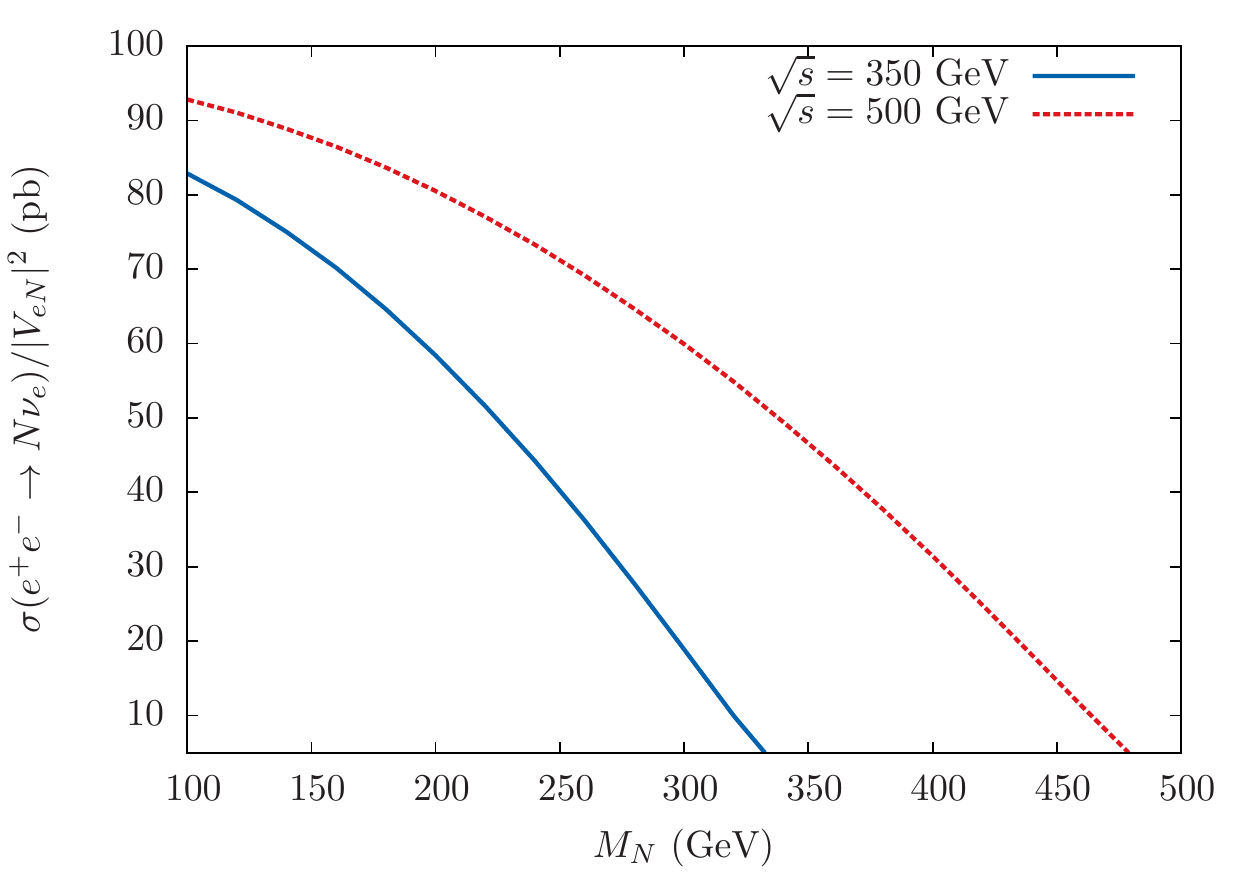} &
\includegraphics[scale=0.6]{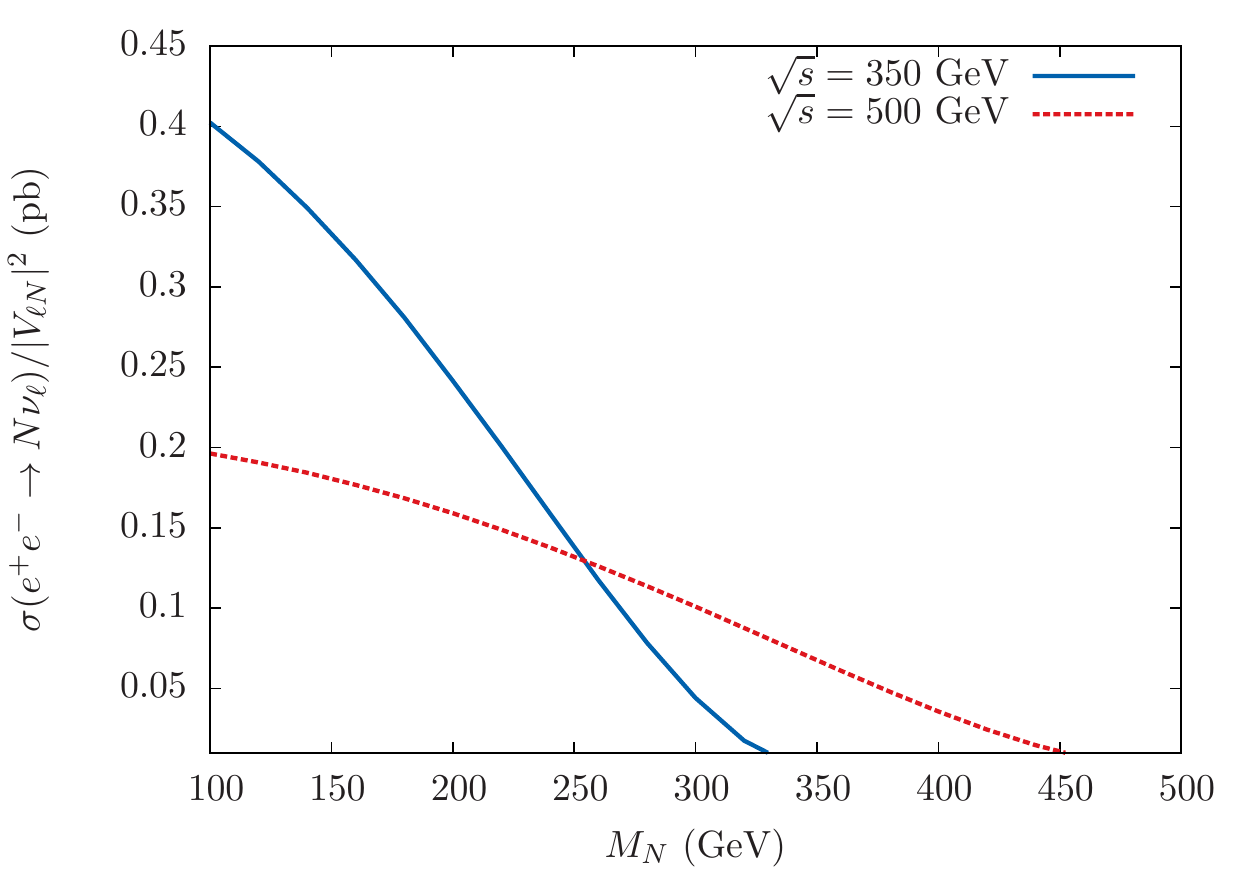} \\
(a) & (b)
\end{tabular}}
\caption{\label{fig:CS} Normalized heavy neutrino production cross section for the channel $ e^{+} e^{-} \to \nu_\ell N$ 
for $\sqrt{s}=350$ and 500 GeV. The left panel corresponds to $\ell=e$ (both $s$ and $t$ channels), whereas the right panel corresponds to $\ell=\mu,\tau$ (only the $s$ channel).}
\end{figure}

A single heavy neutrino of any flavor can be produced through an $s$-channel $Z$-exchange in $e^+e^-$ collisions, as shown in Figure~\ref{fig:FD} (a).\footnote{If $N$ is Majorana, both $\nu_\ell$ and $\bar{\nu}_\ell$ are allowed. However, for brevity, we denote both as simply $\nu_\ell$ in the following.} The corresponding cross section will be proportional to $|V_{\ell N}|^2$. For $\ell=e$, there is an additional $t$-channel $W$-exchange  process, as shown in Figure~\ref{fig:FD} (b). The differential cross section for this process can be found in~\cite{Buchmuller:1991tu, Azuelos:1993qu}. The numerical values of the total production cross sections are shown in Figure~\ref{fig:CS} for two different values of $\sqrt s=350$ and 500 GeV. The left panel of Figure~\ref{fig:CS} corresponds to the case $\ell=e$ and the right panel is for the case $\ell=\mu,\tau$. 

The enhanced cross section in the electron channel is due to the additional $t$-channel contribution. Similarly, for the heavy neutrino decay, $N \to e W$ has the largest branching ratio, as evident from Figure~\ref{fig:BR}. Thus, the channel $e^+e^-\to N\nu_e$, followed by $N\to eW$ and $W\to$ jets, has the largest signal cross section, and hence, can act as the most promising  channel for the discovery of  heavy neutrino mixing with electron neutrinos at a lepton collider. As a matter of fact, the existing direct search limit from LEP~\cite{Achard:2001qv} comes from this channel. 

Due to the relative smallness of the $s$-channel contribution, this production mode does not provide a very promising signal for probing  $|V_{\ell N}|^2$ with $\ell=\mu,\tau$. Also note that the $s$-channel contribution to the cross section decreases with increasing $\sqrt s$, except when the heavy neutrino mass is close to the kinematic threshold, where the phase space suppression becomes more dominant. This is evident from Figure~\ref{fig:CS} (right panel). Using a suitable beam polarization might increase the signal sensitivity for $\ell=\mu,\tau$~\cite{delAguila:2005mf}, but this is of limited practical interest for a small $V_{eN}$.

We note here that since the hadronic activity at an $e^+e^-$ collider is very limited, 
the background can be easily controlled  with the help of simple kinematic cuts. Therefore, the other decay channels of the heavy neutrino, i.e. $N\to Z \nu_\ell,~H \nu_\ell$, can also be used as complementary search channels, even though the corresponding BR are about a factor of two smaller than the $N\to W\ell$ mode.  This will be illustrated in Section~\ref{sec:4}. 

\subsection{$\bm{e^+e^-\to N\ell^\pm W^\mp}$} \label{sec:2.2}
\begin{figure}[t!]
\centering
\subfloat{
\begin{tabular}{ccc}
\includegraphics[scale=0.55]{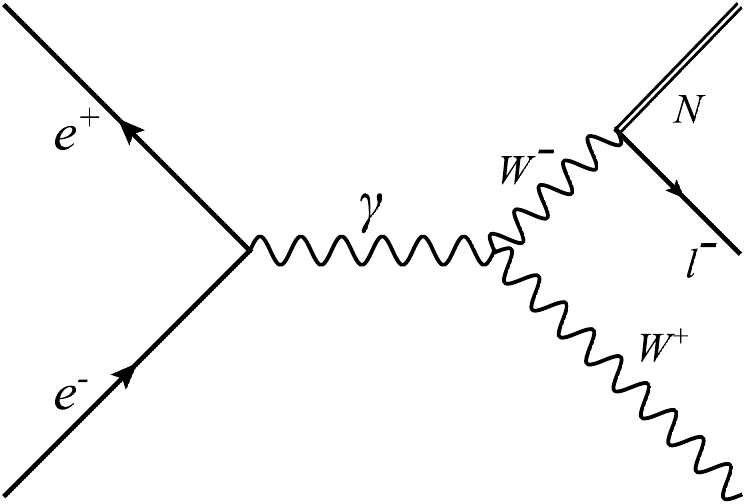} & 
\includegraphics[scale=0.55]{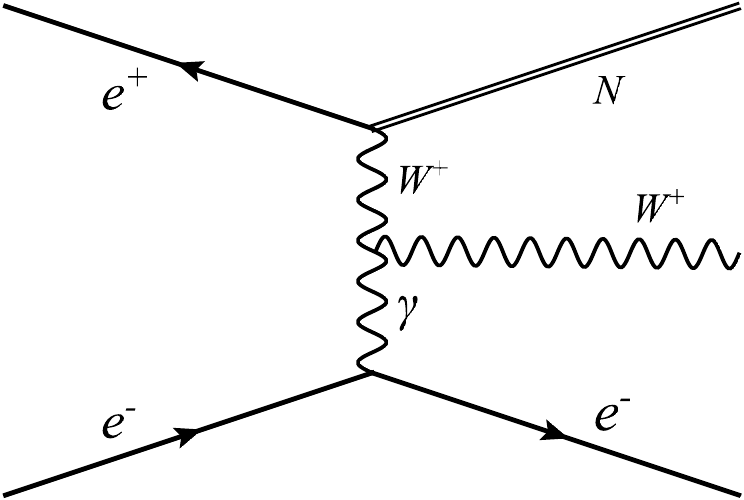} & 
\includegraphics[scale=0.55]{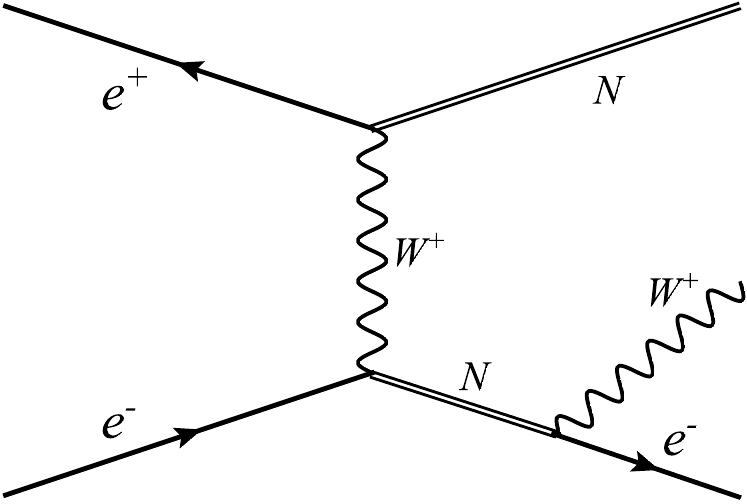} \\
(a) & (b) & (c) 
\end{tabular}}
\caption{\label{fig:FDeNw} Some sample Feynman diagrams for the process $e^+e^-\to N\ell^-W^+$, as discussed in Section~\ref{sec:2.2}. The diagrams of type (a) can probe light-heavy neutrino mixing in all flavors, whereas the diagrams of type (b) and (c) can probe only the electron-sector mixing.}
\end{figure}

The signal discussed in Section~\ref{sec:2.1} is insensitive to the Majorana nature of the heavy neutrino, and hence, does not probe LNV at a lepton collider. Here we present a 
{\em new} production mechanism $e^+e^-\to N\ell^\pm W^\mp$ that can probe the Majorana nature of the heavy neutrinos at an 
$e^+e^-$ collider. This is due to the fact that for a Majorana neutrino, both $\ell^\pm W^\mp$ decay modes are allowed; hence, we get a LNV same-sign dilepton signal, along with two $W$s. Thus, the dominant signal in this case will be $e^\pm \ell^\pm+4j$ which probes LNV. In addition, for $\ell\neq e$, one can in principle also probe the LFV at future colliders.  

\begin{figure}[t!]
\centering
\subfloat{
\begin{tabular}{cc}
\includegraphics[scale=0.55]{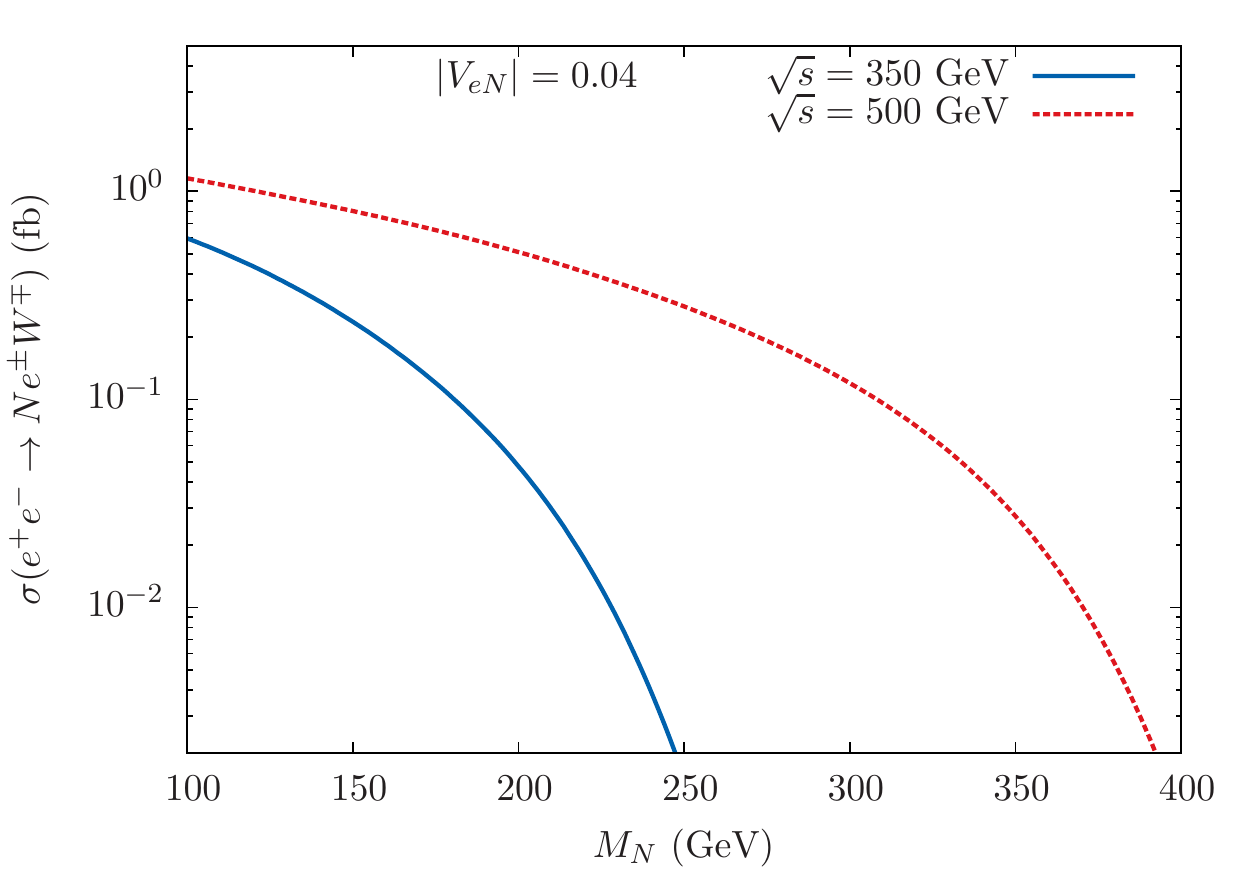} &
\includegraphics[scale=0.55]{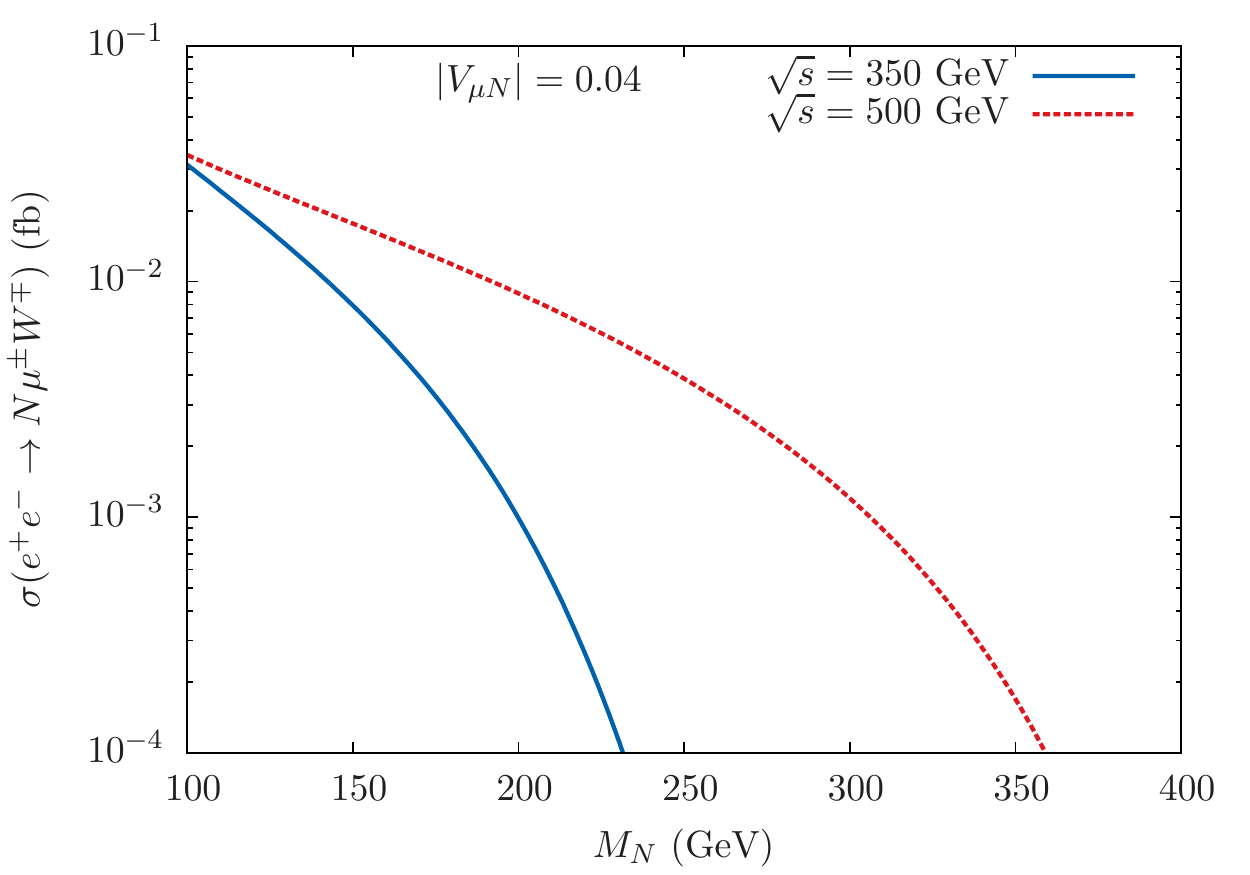} \\
(a) & (b)
\end{tabular}}
\caption{\label{fig:CSeNw} Heavy neutrino production cross section for the channel $e^+e^-\to N\ell^\pm W^\mp$ for $\sqrt{s}=350$ and 500 GeV, and with $|V_{\ell N}|=0.04$. The left panel is for mixing with electrons, whereas the right panel is for mixing with muon and tau sectors.}
\end{figure}

This process gets contribution from several diagrams, and three typical diagrams are shown in Figure~\ref{fig:FDeNw}. For the electron sector, there is a dominant contribution from the $t$-channel photon diagrams [cf. Figure~\ref{fig:FDeNw} (b)], thus leading to an infrared enhancement effect.\footnote{This is similar to the infrared enhancement effects in hadron collisions: $pp\to N\ell^\pm jj$ at the LHC~\cite{Dev:2013wba, Das:2014jxa, Alva:2014gxa}.} This enhancement effect is absent in muon and tau sectors for an $e^+e^-$ collider. The total production cross section for this process is given in Figure~\ref{fig:CSeNw}. The left panel shows the cross section for electron-sector mixing and the right panel is for $\ell=\mu,\tau$. Note that this process involves diagrams with a heavy neutrino as as intermediate state [cf. Figure~\ref{fig:FDeNw} (c)], and therefore, the total cross section for this process does not simply scale as $|V_{\ell N}|^2$, since the total heavy neutrino decay width in the propagator explicitly depends on $|V_{\ell N}|^2$.  The cross section values shown in Figure~\ref{fig:CSeNw} were obtained assuming a typical mixing parameter $|V_{\ell N}|=0.04$. A detailed detector-level simulation is done in Section~\ref{sec:3} to derive the sensitivity for this process at future lepton colliders. 

\begin{figure}[t!]
\centering
\includegraphics[scale=0.55]{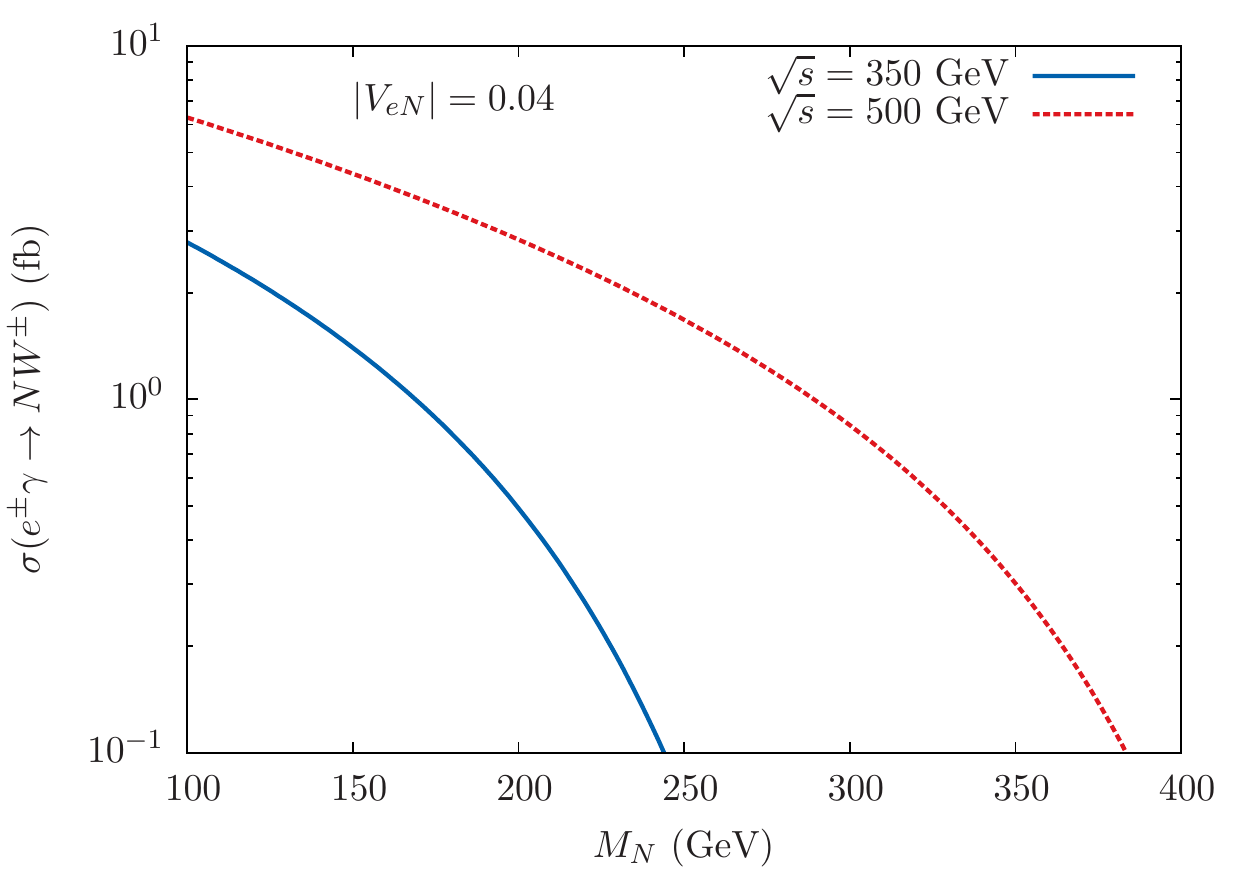}
\caption{Heavy neutrino production cross section for the channel $e^\pm \gamma \to N W^\pm$, with the real photon coming from one of the electron or positron beams, for $\sqrt{s}=350$ and 500 GeV, and with $|V_{e N}|=0.04$.}
\label{fig:CSeaNW}
\end{figure}

A related process to consider is $e^\pm \gamma \to N W^\pm$, where the real photon comes from one of the electron or positron beams. This process can be calculated using the Weizsacker-Williams equivalent photon approximation (EPA)~\cite{Weizsacker:1934sx, Williams:1934ad}. The results are shown in Figure~\ref{fig:CSeaNW}. Again in this case, one can have a same-sign dilepton signal, when $N\to \ell^\pm W^\mp$ and the other $W$ with the same sign as the lepton in the final state decays leptonically.  

\subsection{$\bm {e^+e^-\to Ne^\pm \ell^\mp \nu_\ell}$}\label{sec:2.3}
\begin{figure}[t!]
\centering
\begin{tabular}{cc}
\includegraphics[scale=0.55]{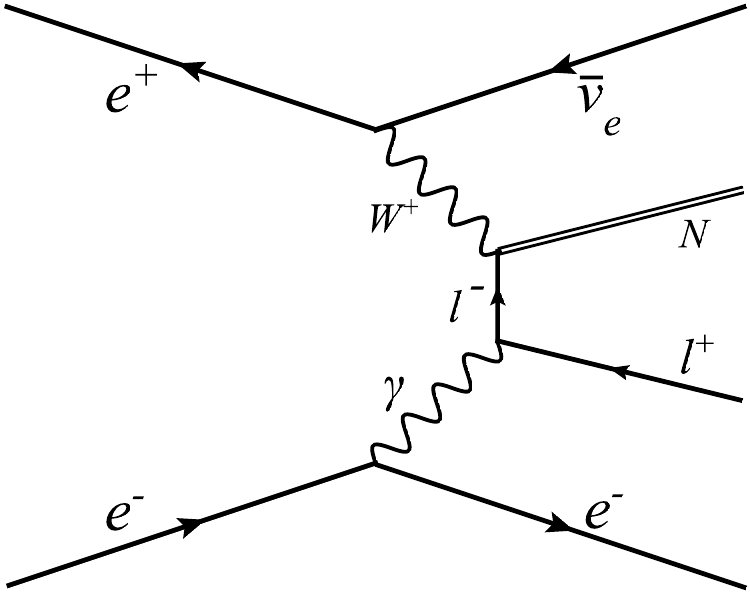} & \hspace{1cm}
\includegraphics[scale=0.55]{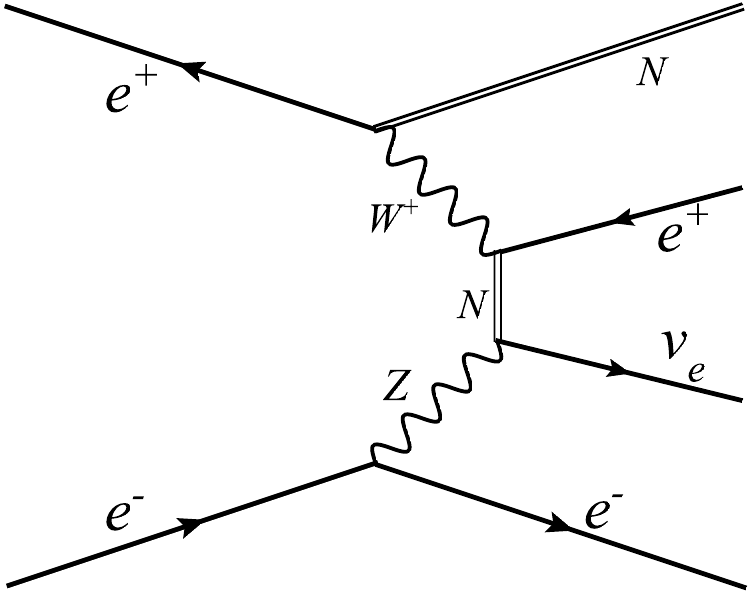} \\
(a) & \hspace{1cm} (b)
\end{tabular}
\caption{Sample Feynman diagrams for the process $e^+e^-\to Ne^\pm \ell^\mp \nu_\ell$, as discussed in Section~\ref{sec:2.3}.}
\label{fig:FDee2Neev}
\end{figure}

\begin{figure}[t!]
\centering
\subfloat{
\begin{tabular}{cc}
\includegraphics[scale=0.55]{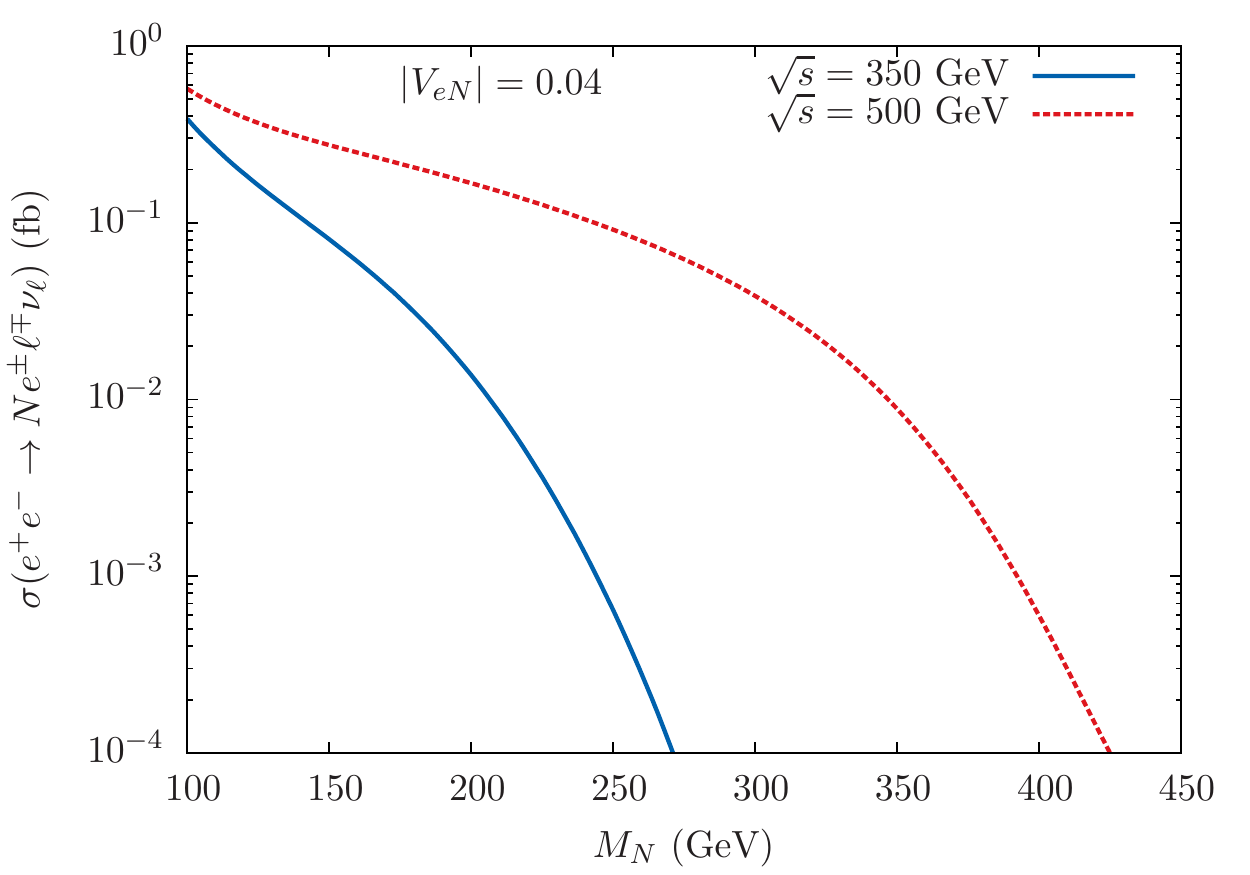} &
\includegraphics[scale=0.55]{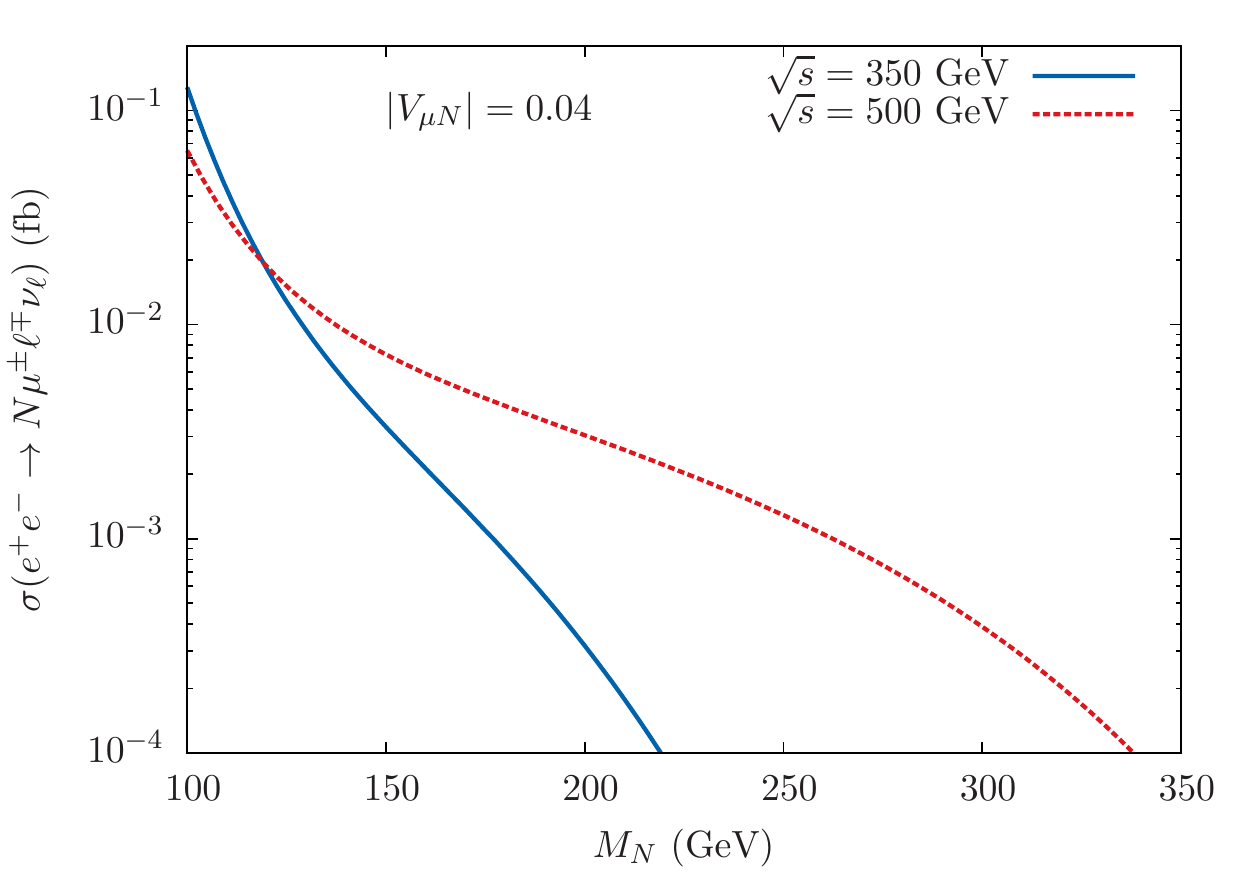} \\
(a) & (b)
\end{tabular}}
\caption{Heavy neutrino production cross section for the channel $e^+e^-\to N \ell^\pm \ell'^\mp \nu_\ell$ for $\sqrt{s}=350$ and 500 GeV, and with mixing $|V_{\ell N}|=0.04$. The left panel corresponds to $\ell=e$ and the right panel is for $\ell=\mu$, whereas $\ell'$ could be any of the three lepton flavors.}
\label{fig:CSee2N2v}
\end{figure}

Similar to the case discussed in Section~\ref{sec:2.2}, the process $e^+e^-\to Ne^\pm \ell^\mp \nu_\ell$ also contains infrared enhanced $t$-channel contributions mediated by virtual photons, as illustrated in Figure~\ref{fig:FDee2Neev}(a). The cross section for this process is given  in Figure~\ref{fig:CSee2N2v}. Similar to the situation in Section~\ref{sec:2.2}, this process involves Feynman diagrams with heavy neutrino propagators [cf.~Figure~\ref{fig:FDee2Neev}(b)], and hence, the cross section does not simply scale as $|V_{\ell N}|^2$. 

\begin{figure}[t!]
\centering
\includegraphics[scale=0.55]{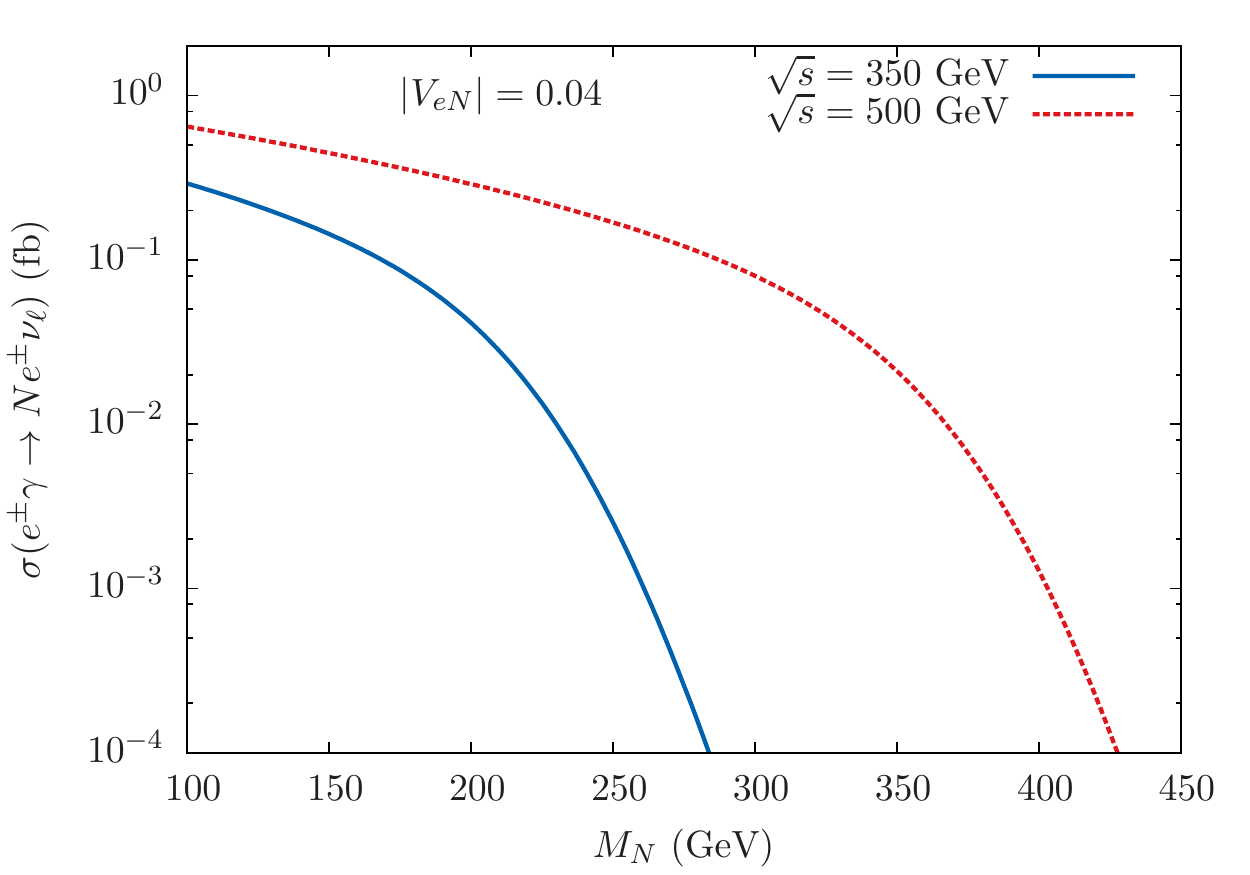}
\caption{Heavy neutrino production cross section for the channel $e^\pm \gamma \to N \ell^\pm \nu_\ell$, with the real photon coming from one of the electron or positron beams, for $\sqrt{s}=350$ and 500 GeV, and with $|V_{e N}|=0.04$.}
\label{fig:CSea2Nev}
\end{figure}

A related process to consider here is $e^\pm \gamma \to N \ell^\pm \nu_\ell$, where the real photon comes from one of the electron or positron beams. We calculate the cross section for this process using EPA and the results are shown in Figure~\ref{fig:CSea2Nev}. 

\subsection{$\bm{e^+e^-\to ZH\to ZN\nu_\ell}$}\label{sec:2.4}
\begin{figure}[t!]
\centering
\includegraphics[scale=0.55]{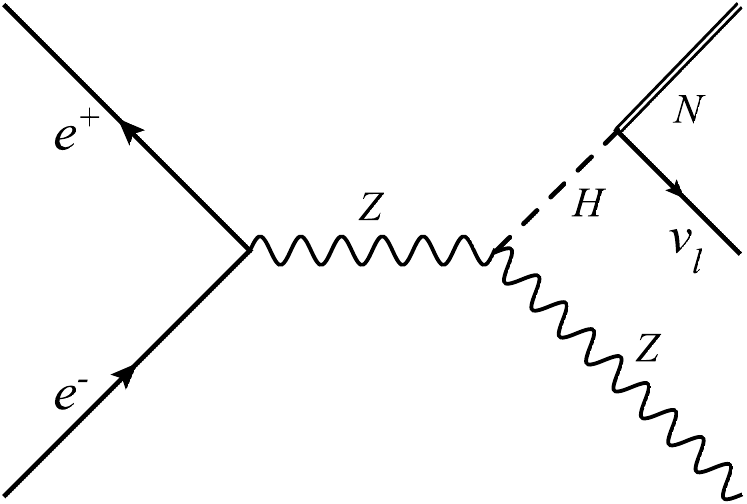}
\caption{Feynman diagram for the process $e^+e^-\to ZH\to ZN\nu_\ell$.}
\label{fig:FDZH}
\end{figure}

\begin{figure}[t!]
\centering
\includegraphics[scale=0.55]{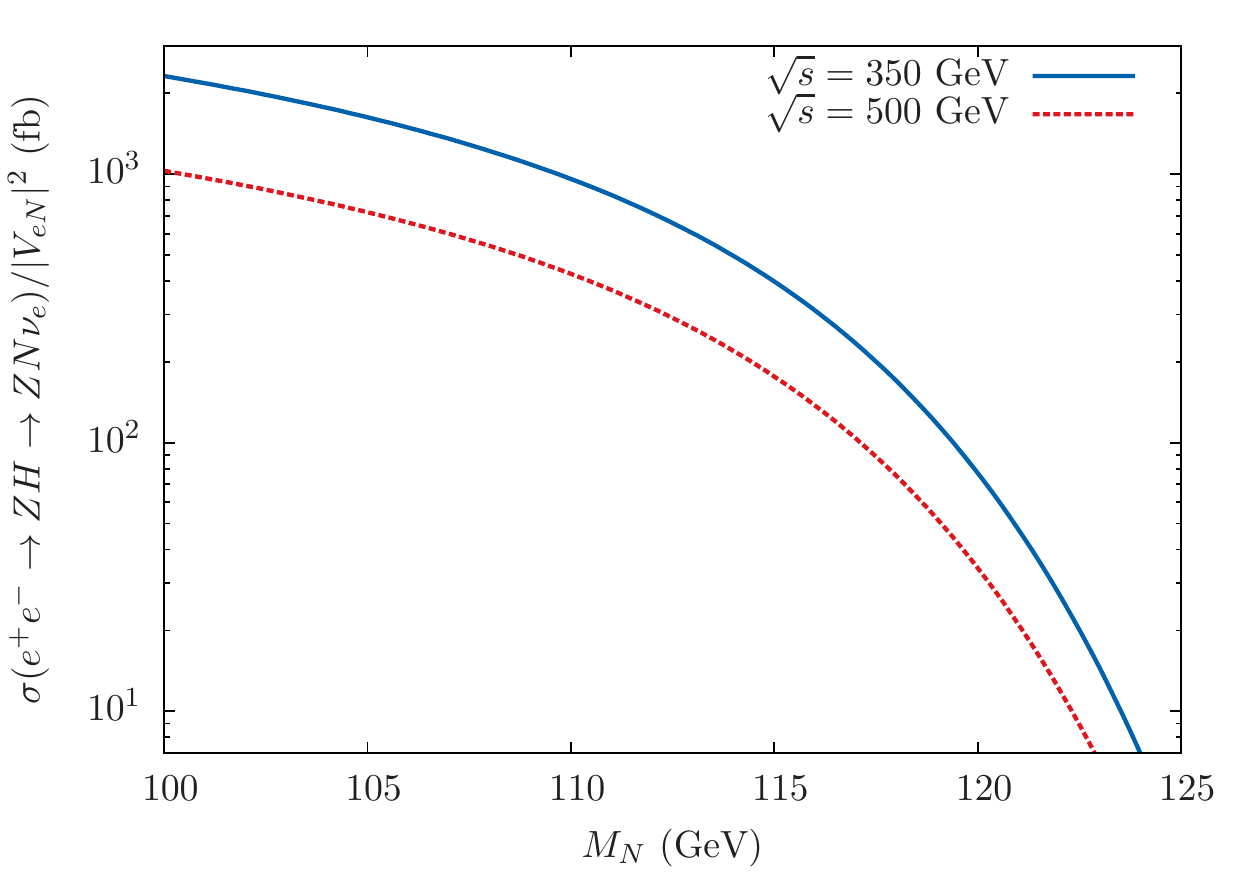}
\caption{Normalized heavy neutrino production cross section for the channel $e^+e^-\to ZN\nu_\ell$ for $\sqrt{s}=350$ and 500 GeV.}
\label{fig:CSZH}
\end{figure}

This is the Higgsstrahlung process, followed by the decay of Higgs to $N\nu_\ell$, as shown in Figure~\ref{fig:FDZH}. However, this process is only effective for $M_N$ values below the Higgs mass, as evident from the cross section values shown in Figure~\ref{fig:CSZH}. This process can directly probe the heavy neutrino mixing with all neutrino flavors. Note that for the heavy neutrino mass in the vicinity of the Higgs mass, the Higgs decay width will get significantly modified due to the new decay channels available, and using the Higgs signal strength data, one 
can also derive indirect constraints on the mixing parameter~\cite{BhupalDev:2012zg, Cely:2012bz}.  

\subsection{$\bm {e^-e^- \to W^-W^-}$}\label{sec:2.5}

\begin{figure}[!h]
\centering
\includegraphics[scale=0.55]{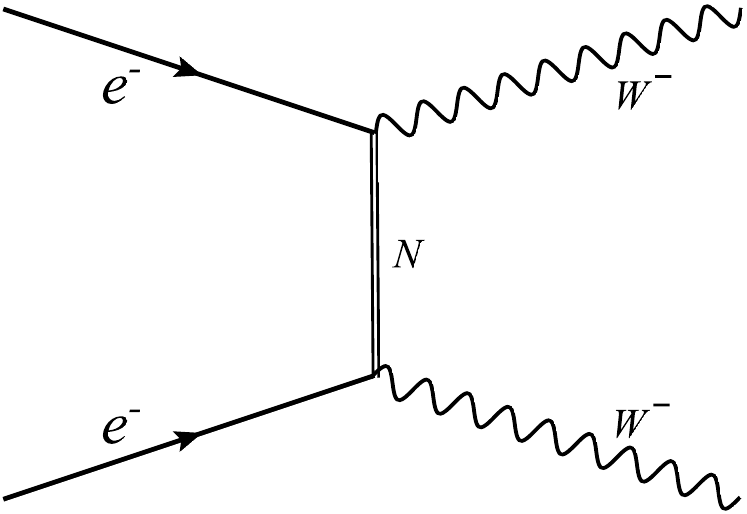}
\caption{Feynman diagram for the $t$-channel $N$ exchange.}
\label{fig:ee2ww}
\end{figure}

Experimentally, it is possible to switch from $e^+e^-$ mode to the $e^-e^-$ mode within the same experimental setup~\cite{Adolphsen:2013kya}. In this case, the Majorana nature  of the heavy neutrino can be tested by observing the LNV process
 $e^-e^-\to W^-W^-$, as shown in Figure~\ref{fig:ee2ww}. This process is mediated by a $t$-channel $N$ exchange,\footnote{A similar diagram exists with a $t$-channel light Majorana neutrino exchange. However, its contribution to the cross section is extremely small, suppressed by the light neutrino mass.}
and is free from SM background, except in the case when two additional leptons are produced and lost in the beam pipe. This can be considered as the lepton-collider analogue of the $0\nu\beta\beta$ process. Moreover, for a linear collider like the ILC, it is also possible to use polarized electron beams. By using two left-handed electron beams, the cross-section for the $e^-e^-\to W^-W^-$ channel will 
increase by a factor of four as compared to two unpolarized beams. On the other hand, for a circular collider like the FCC it is not possible to
retain the polarization of the colliding beams due to the strong magnetic field used to bend the beam tracks. 

\begin{figure}
\centering
\includegraphics[scale=0.55]{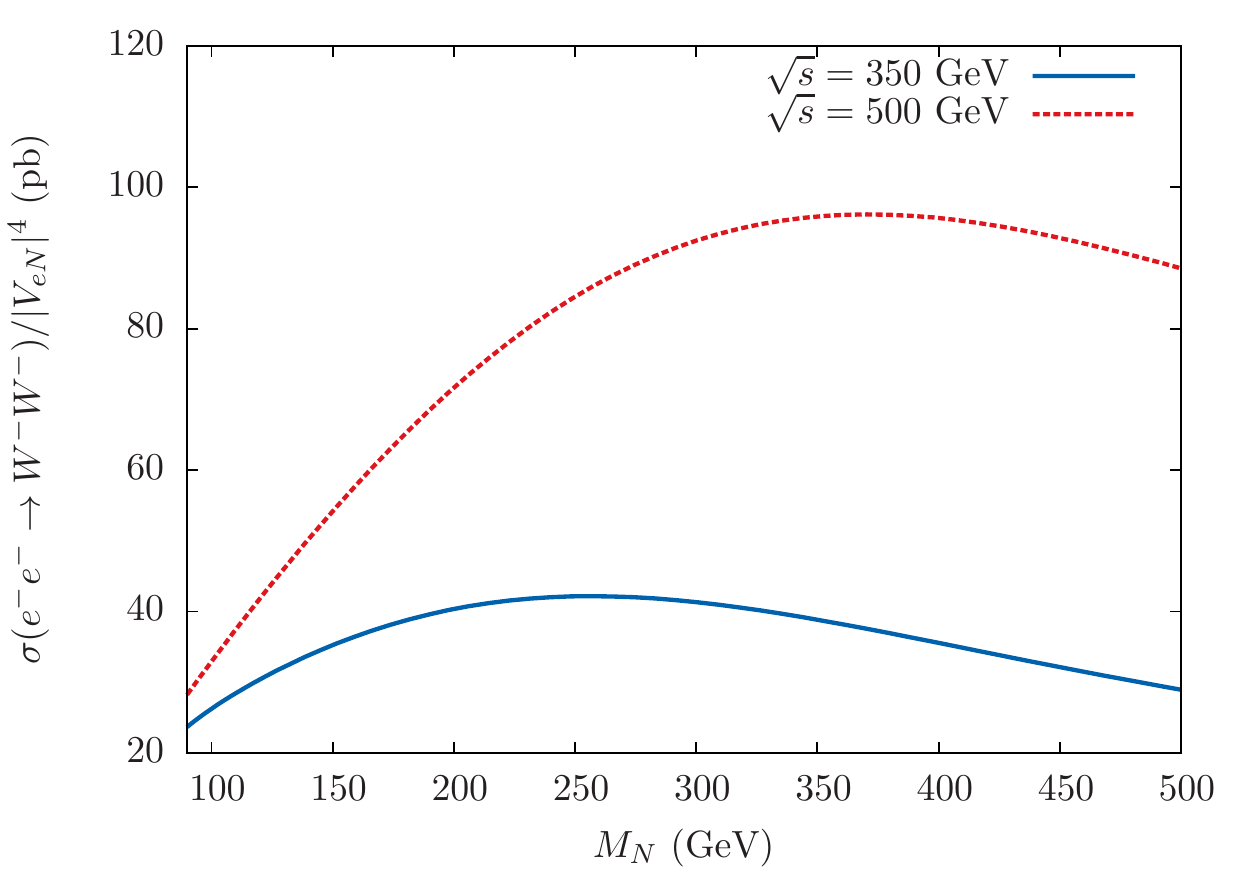}
\caption{Normalized cross section of the process $e^{-} e^{-} \to W^{-} W^{-}$  for $\sqrt{s}=350$ and  500 GeV. }\label{fig:CSww}
\end{figure}

The cross section for the channel shown in Figure~\ref{fig:ee2ww} is proportional to  $|V_{eN}|^4$ and the normalized cross section is shown in Figure~\ref{fig:CSww} for $\sqrt s=350$ and 500 GeV. The produced $W$s can decay either hadronically or leptonically, thus generating the final states with $4j$, $2\ell^{-}+ \slashed{E}$ or $2j+l^{-} + \slashed{E}$. The cross section values after the selection cuts and the signal sensitivity will be discussed in Section~\ref{sec:3}.

\section{Cut-based Analysis}\label{sec:3}

We have used the MC event generator \textsc{MadGraph5}~\cite{Alwall:2011uj} to  generate all signal and background events. The subsequent parton showering and hadronization have been done using \textsc{Pythia6}~\cite{Sjostrand:2006za}. We have used \textsc{Delphes3}~\cite{deFavereau:2013fsa} for simplified detector simulation relevant for a generic $e^+e^-$ collider. In this section, we present a cut-based analysis (CBA) to discriminate signal from the background and to derive the  sensitivity limits for heavy neutrinos. In Section~\ref{sec:4}, we have employed a more sophisticated MVA for better signal-to-background discrimination, which leads to a better significance. 
 
For our subsequent collider analysis, we apply the following basic trigger cuts, in addition to those in Eq.~\eqref{lepton_TC}, to identify leptons and jets in the final state: 
\begin{align}
 p_T^{j} \ > \ 20~{\rm GeV}, \qquad |\eta^j| \ < \ 2.5, \qquad  \Delta R^{jj} \ > \ 0.4, \qquad \Delta R^{\ell j} \ > \ 0.4 \; .
\label{jet_TC}
\end{align} 
Jets are clustered using the anti-$k_T$ algorithm~\cite{Cacciari:2008gp} implemented in \textsc{FastJet}~\cite{Cacciari:2011ma} with clustering parameter $R=0.4$. Then we apply further selection cuts for each signal discussed below, depending on its kinematics, in order to isolate it from the SM background, if present.

In the following, we consider various final states corresponding to the different heavy neutrino production modes, as discussed in Section~\ref{sec:2}. Since the cross sections involving the mixing with the muon and tau sector are much smaller compared to the electron sector, we only consider the latter to derive the discovery prospects of the heavy neutrino at a future lepton collider. 

\subsection{$\bm{e^+e^-\to N\nu_e\to e+2j+\slashed{E}}$}
\label{subsec:ee2Nv}

The dominant SM background for this final state comes from the $e^+e^-\to W^+W^-$ channel, followed by the leptonic decay of one $W$ and the hadronic decay of the other. Therefore, it is expected that the invariant mass of the electron and the neutrino, viz. $M(e,\slashed{E})$ peaks around the $W$-mass for the background, while for the signal, one does not expect such a behavior. On the other hand, the invariant mass distribution of the isolated electron and the two hardest tagged jets, $M(e,j_1,j_2)$, shows a peak near $M_N$ for the signal, but not for the background. These two distributions for the signal and background are shown later in Figure~\ref{fig:varBDT} in the context of a MVA. Therefore, $M(e,\slashed{E})$ and $M(e,j_1,j_2)$ distributions can be used to separate the signal from the background in order to get a better sensitivity. Based on this observation, we demand the following selection cuts specific to this signal: exactly one isolated electron and at least two tagged jets satisfying the basic trigger cuts, no photon with $p_T > 20$ GeV and 
\begin{align}
|M(e,j_1,j_2) - M_{N}| \ < \ 40~{\rm GeV}, \qquad |M(e,\slashed{E}) - M_W| \ > \ 20~\textrm{GeV} \; .
\label{eq:SCCBA}
\end{align}

\begin{table}[!h]
\centering	
\begin{tabular}{|c|c|c|c|c|c|}
\hline 
$M_N$ & \multicolumn{2}{|c|}{Signal (fb)} & \multicolumn{2}{|c|}{Background (fb)} & $\mathcal{N_S}$ \\ \cline{2-5} 
 (GeV) & TC & SC & TC & SC & $\overline{\sqrt{\mathcal{N_S}+\mathcal{N_B}}}$ \\ 
\hline 
250 & 21.94 & 18.96 & 638.25 & 68.99 & 20.22 \\ 
\hline 
350 & 16.33 & 13.48 & 638.25 & 46.11 & 17.46 \\ 
\hline 
\end{tabular} 
\caption{Signal ($\sigma_\mathcal{S}$) and background ($\sigma_\mathcal{B}$) cross sections after the trigger cuts (TC) and the selection cuts (SC) as defined in the text for different $M_N$ with $|V_{eN}|=0.04$ at $\sqrt{s}=500$ GeV. The significance is computed using 100 fb$^{-1}$ integrated luminosity.}
\label{tab:CSac}
\end{table}

\begin{figure}[t!]
\centering
\subfloat{
\begin{tabular}{cc}
\includegraphics[scale=0.6]{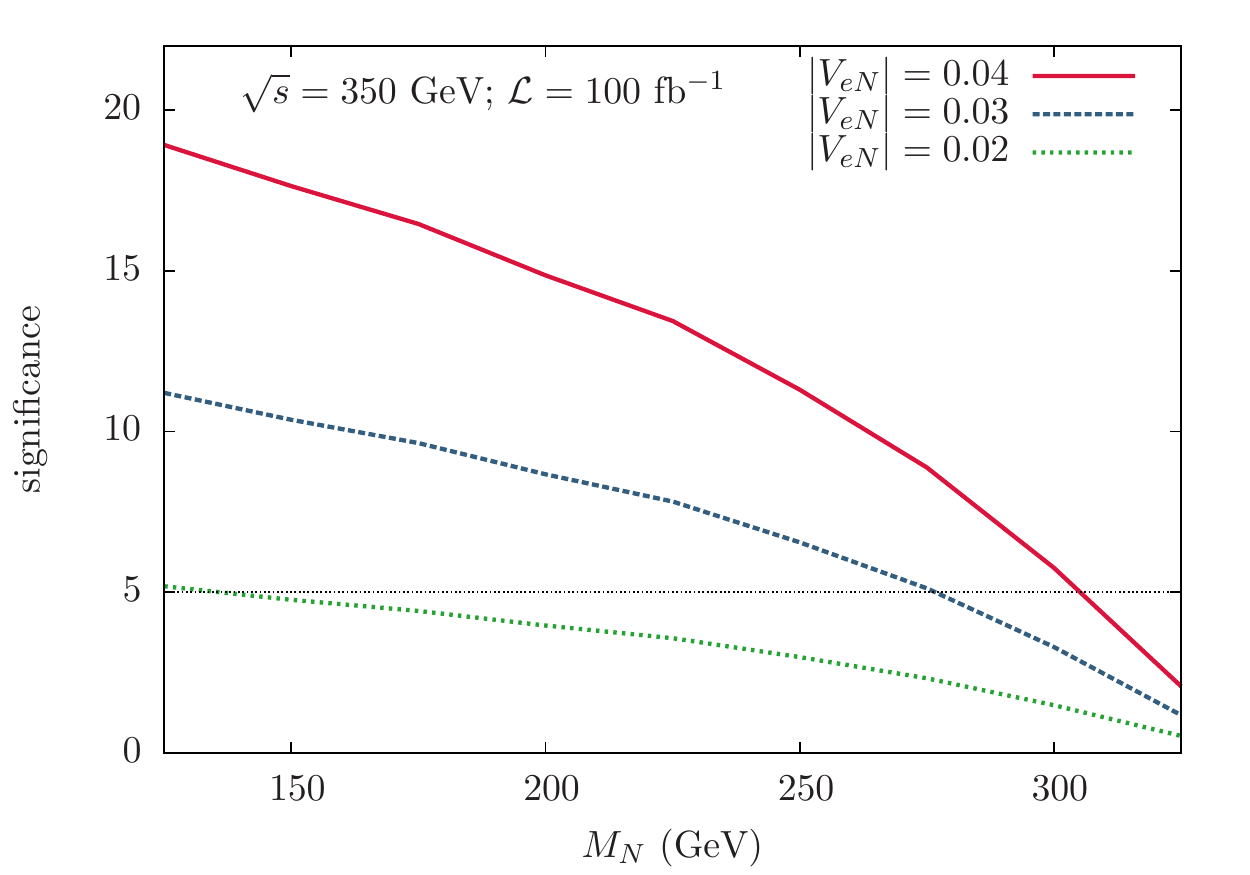} &
\includegraphics[scale=0.6]{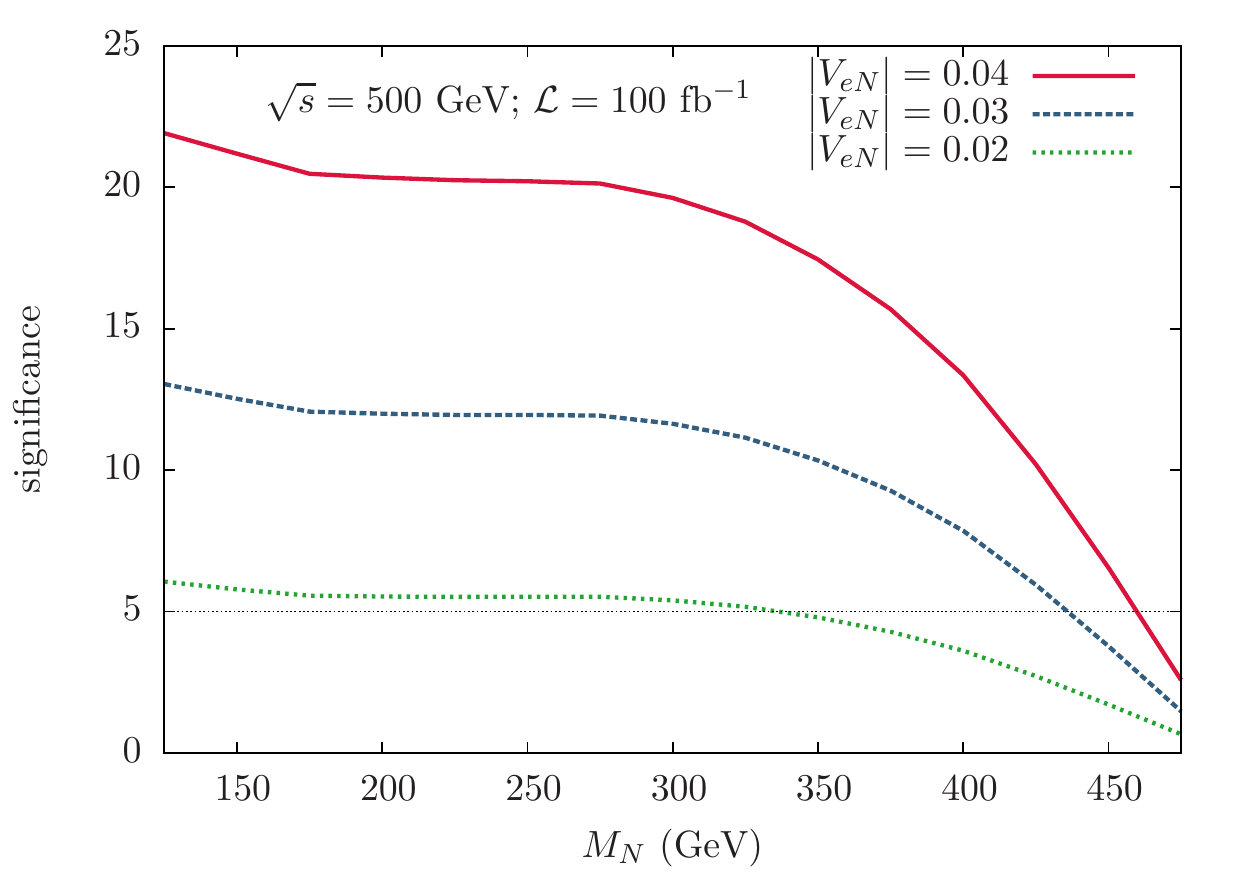} \\
(a)&(b)
\end{tabular}}
\caption{Significance defined in Eq.~\eqref{signi} as functions of $M_N$ for different mixing values at (a) $\sqrt{s}=350$ GeV and (b) $\sqrt{s}=500$ GeV, with $\mathcal{L}=100$ fb$^{-1}$.}
\label{fig:signi}
\end{figure}

After imposing these selection criteria and including the detector effects, we compute the signal and background cross sections for two illustrative values of the heavy neutrino mass $M_N=250,~350$ GeV at $\sqrt s=500$ GeV and for the mixing parameter $|V_{eN}|=0.04$. The results are shown in Table~\ref{tab:CSac}. The signal significance 
\begin{align}
n \ = \ \frac{\mathcal{N_S}}{\sqrt{\mathcal{N_S}+\mathcal{N_B}}} \; ,
\label{signi}
\end{align}
where $\mathcal{N_S}$ and $\mathcal{N_B}$ are the number of signal and background events respectively, has been computed with an integrated luminosity of 100 fb$^{-1}$.

In Figure~\ref{fig:signi}, we show the significance $n$ defined by Eq.~\eqref{signi} as a function of the heavy neutrino mass $M_N$, for different mixing values $|V_{eN}|=0.04, 0.03$ and $0.02$ at $\sqrt{s}=350$ and 500 GeV for $\mathcal{L} = 100$ fb$^{-1}$. We see that mixing values up to $|V_{eN}|=0.02$ can be probed with a significance of $\gtrsim 5\sigma$ for heavy neutrino mass $M_N\lesssim 350$ GeV at $\sqrt{s} = 500$ GeV. 

\begin{figure}[t!]
\centering
\includegraphics[scale=0.75]{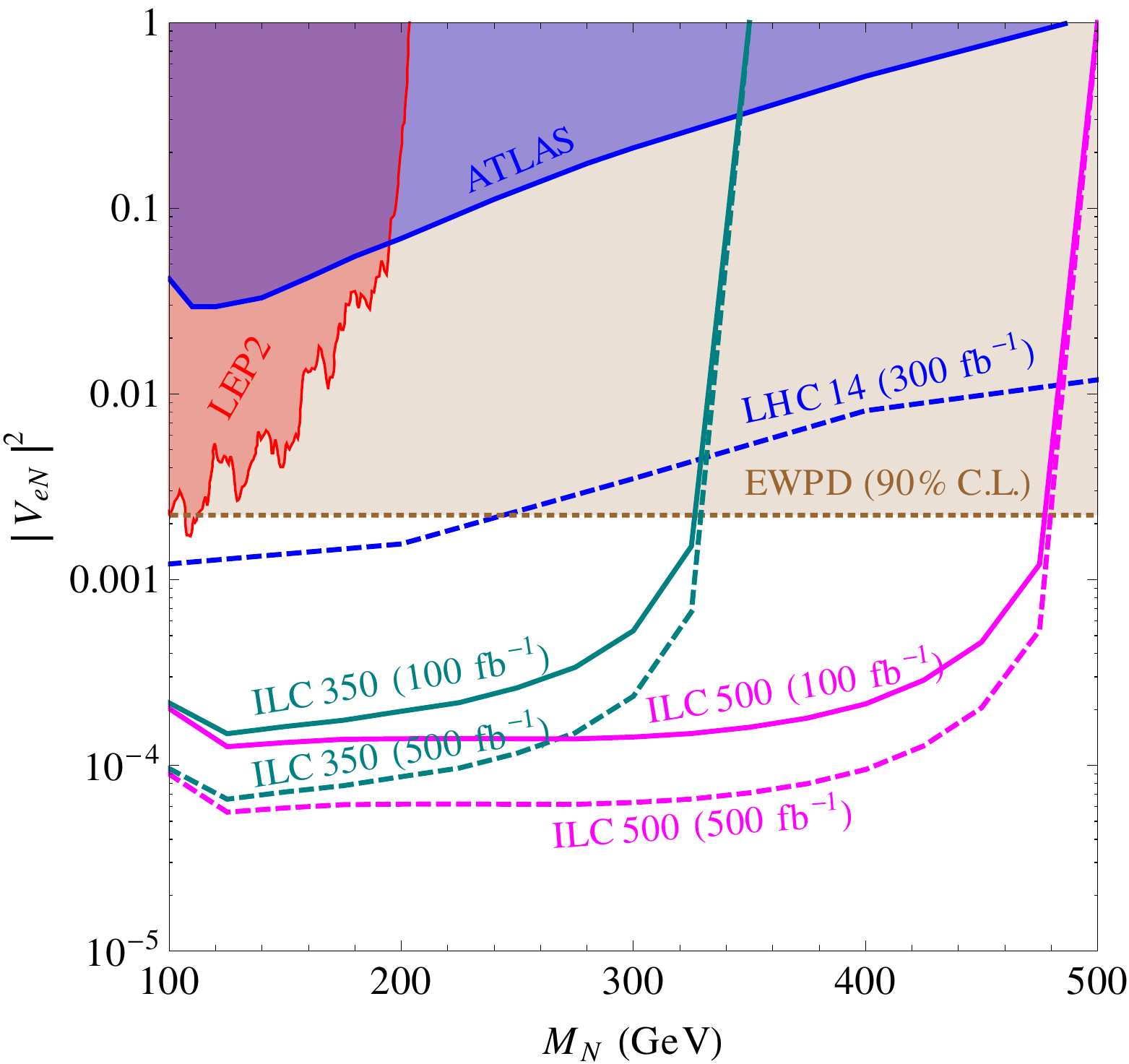}
\caption{The projected $95\%$ C.L. ILC sensitivity of the light-heavy neutrino mixing parameter $|V_{eN}|^2$ as a function of the heavy neutrino mass $M_N$ for $\sqrt s=350$ (green) and 500 GeV (magenta), and with 100 fb$^{-1}$ (solid) and 500 fb$^{-1}$ (dashed) integrated luminosity. Also shown are the current 90\% C.L. indirect exclusion limit from electroweak precision data (EWPD; brown, dotted)~\cite{Antusch:2015mia}, 95\% C.L. direct exclusion limits from LEP 2 (red, shaded)~\cite{Achard:2001qv} and $\sqrt s=8$ TeV LHC data (blue, shaded)~\cite{klinger:2014}, as well as a projected 95\% C.L. limit from $\sqrt s=14$ TeV LHC with 300 fb$^{-1}$ integrated luminosity (blue, dashed)~\cite{Deppisch:2015qwa}.} \label{fig:sensitivity}
\end{figure}

In order to make a comparative study of the projected sensitivity obtained here with the existing and future limits from other contemporary experiments, we show the 95\% C.L. projected ILC limits on $|V_{eN}|^2$ as a function of $M_N$ in Figure~\ref{fig:sensitivity}. We have considered two configurations with $\sqrt s=350$ (green curves) and 500 GeV (magenta curves) and with integrated luminosities ${\cal L}=100~{\rm fb}^{-1}$ (solid) and $500~{\rm fb}^{-1}$ (dashed). For larger $M_N$, the cross section decreases, as evident from Figure~\ref{fig:CS}, thus leading to a weaker limit on $|V_{eN}|^2$. For comparison, we have also shown the current 95\% C.L. direct exclusion limits from LEP 2 (red, shaded region)~\cite{Achard:2001qv} and $\sqrt s=8$ TeV LHC data (blue, shaded region)~\cite{klinger:2014}, as well as a projected 95\% C.L. limit from $\sqrt s=14$ TeV LHC with 300 fb$^{-1}$ integrated luminosity (blue, dashed curve)~\cite{Deppisch:2015qwa}. The best current limit is the indirect limit at 90\% C.L. (brown, dotted line) derived from an analysis of the EWPD~\cite{Antusch:2015mia}.\footnote{For a detailed discussion of the present bounds on light-heavy neutrino mixing, see~\cite{Deppisch:2015qwa}.} As can be seen from Figure~\ref{fig:sensitivity}, a linear collider can significantly improve the heavy neutrino sensitivity up to mass values close to its kinematic threshold. Note that here we have not shown the constraints from $0\nu\beta\beta$ (see e.g.~\cite{Ibarra:2010xw, Mitra:2011qr, Nemevsek:2012iq, Dev:2013vxa, Dev:2014xea, Faessler:2014kka}), since these limits are highly model dependent and can be significantly weakened in certain cases due to cancellation between different contributions~\cite{Pascoli:2013fiz}.

As far as the other decay modes of the heavy neutrino are concerned, it can go to $Z\nu$ and $H\nu$ final states, followed by $Z\to \ell^+\ell^-,jj$, 
$H\to b\bar{b}$ and $W\to \ell \nu$ decay modes, thus giving rise to $jj+\slashed{E}$, $\ell^+\ell^-+\slashed{E}$, $b\bar{b}+\slashed{E}$ final states. For illustration, we show in Table~\ref{crosstab} the signal cross sections of these final states for two benchmark values of $M_N$ at $\sqrt s=500$ GeV. Note that the $2\ell+\slashed{E}$ final state has two contributions coming from the leptonic decay of either $W$ or $Z$, as shown by the last two columns of Table~\ref{crosstab}. 
Due to the relatively clean environment at a lepton collider, these processes can still lead to a non-negligible number of events with $500~{\rm fb}^{-1}$ luminosity. Note that for the channel involving Higgs decay, we have taken into account the correct Higgs decay width at next-to-next-to-leading order (NNLO): $\Gamma(H\to b\bar{b})_{\rm NNLO}=0.0023$ GeV~\cite{Heinemeyer:2013tqa}, whereas the corresponding width obtained from \textsc{MadGraph5} at leading order is 0.0053 GeV.

\begin{table}[!h]
\centering	
\begin{tabular}{|c|c|c|c|c|}
\hline 
$M_N$ (GeV) &  $2j +\slashed{E}$ (fb)  &  $2 b+ \slashed{E}$ (fb) & $ 2\ell+ \slashed{E}~ (Z)$ (fb) & $ 2\ell+ \slashed{E}~ (W)$ (fb) \\ 
\hline
250 & 16.41   & 9.03  & 1.641 & 11.05 \\
\hline
400 &   7.415    & 5.20  & 0.7417 & 4.926  \\
\hline
\end{tabular}
\caption{Parton level signal cross sections for the subdominant decay modes of the heavy neutrino produced in  the $e^+e^-\to N\nu_\ell$ process after trigger and selection cuts for illustrative cases of heavy neutrino masses $M_N= 250$ and $ 400$ GeV and for mixing parameter $|V_{\ell N}|=0.04$. }
\label{crosstab}
\end{table}

\subsection{$\bm{e^+e^-\to Ne^{\pm}W^{\mp}\to e^{\pm}e^{\pm}+4j}$}
\label{subsec:ee2NeW}

\begin{figure}[t!]
\centering
\subfloat{
\begin{tabular}{cc}
\includegraphics[scale=0.6]{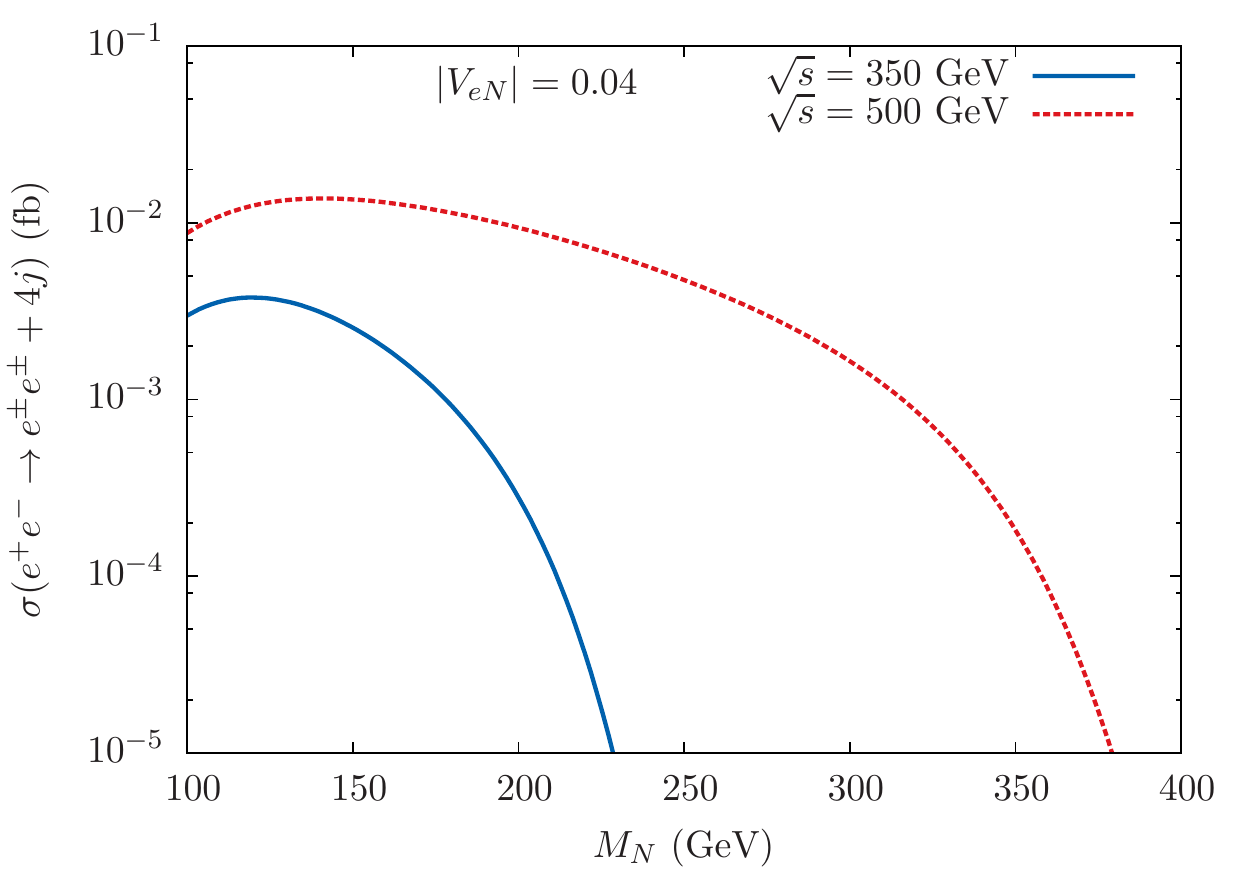} &
\includegraphics[scale=0.6]{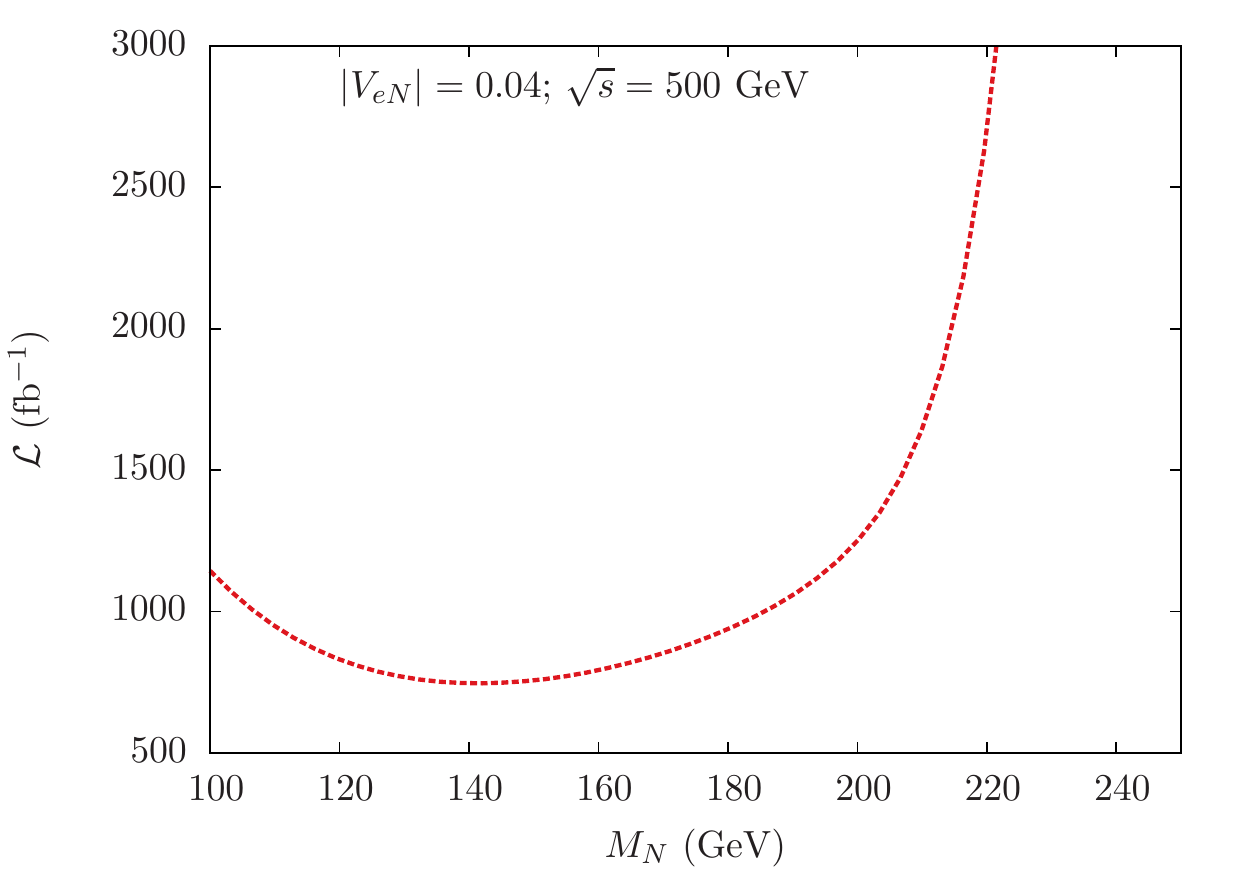} \\
(a)&(b)
\end{tabular}}
\caption{(a) cross sections for the process $e^+e^-\to Ne^{\pm}W^{\mp}\to e^{\pm}e^{\pm}+4j$ after trigger cuts and including detector efficiency at $\sqrt{s}=350,~500$ GeV and (b) the required luminosity to observe 10 signal events with $\sqrt{s}=500$ GeV. Here we have taken the mixing $|V_{eN}|=0.04$.}
\label{fig:same-sign}
\end{figure}

In this subsection, we consider the LNV process of $e^+e^-\to Ne^\pm W^\mp \to e^\pm e^\pm +4j$, as discussed in Section~\ref{sec:2.2}.  
 We demand in the final state, two electrons (or two positrons) and at least four tagged jets satisfying the basic trigger cuts given in Eqs.~\eqref{lepton_TC} and \eqref{jet_TC}. The signal cross section 
after imposing the cuts and including the detector efficiency is given in Figure~\ref{fig:same-sign} (a). The SM background for this process, mostly coming from $4\ell +4j$ final states with two leptons lost in the beam pipe, is negligible $\lesssim {\cal O}(10^{-5}~{\rm fb})$. 
Hence, we do not perform a detailed background simulation for this process.
It is evident from Figure~\ref{fig:same-sign}(a) that  the signal cross section is also small ($< 10^{-2}$ fb) and  drops sharply with increasing $M_N$. This makes it a challenging task to observe a reasonable number of signal events with 100 $\rm{fb}^{-1}$ integrated luminosity. In Figure~\ref{fig:same-sign} (b), we  show the required integrated luminosity for observing 10 signal events. We find that $\mathcal{L}\gtrsim 700$ fb$^{-1}$ is required for this purpose at a $\sqrt{s}=500$ GeV ILC. This is nevertheless within reach of future lepton colliders~\cite{Barletta:2014nka}. Although this channel has a smaller cross section, as compared to the $e+ 2j+\slashed{E}_T$ channel discussed in Section~\ref{subsec:ee2Nv}, an observation of even a few LNV events in this channel will concretely establish the Majorana nature of heavy neutrinos, with profound implications on our understanding of the origin of neutrino mass.     

\subsection{$\bm{e^-e^-\to W^-W^-\to 4j}$}

\begin{figure}[t!]
\centering
\includegraphics[scale=0.55]{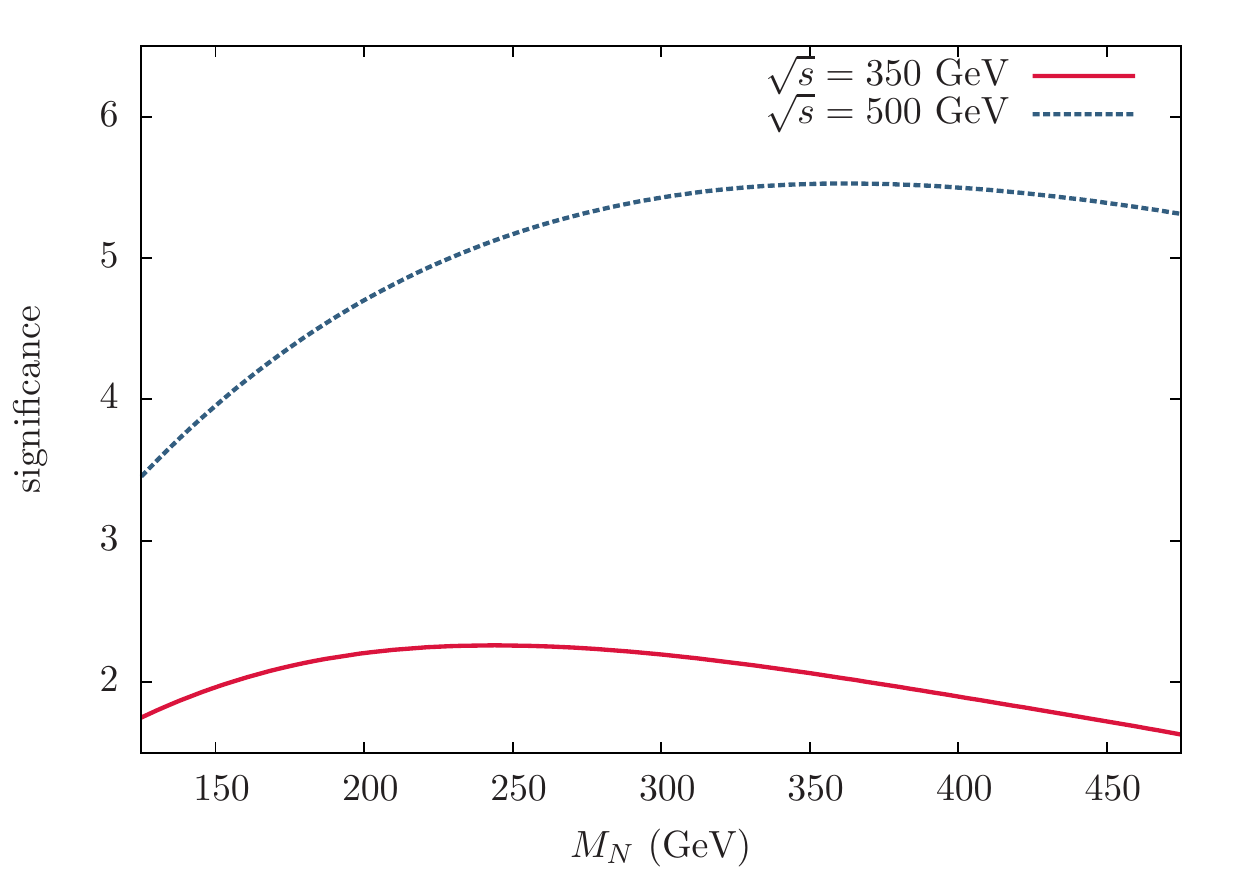}
\caption{The significance as a function of $M_N$ for the $e^-e^-\to 4j$ channel for an integrated luminosity of 500 fb$^{-1}$ at $\sqrt{s}=350,500$ GeV.}
\label{fig:ee24j-sig}
\end{figure}

In $e^+e^-$ colliders, the positron beam can be readily switched to an electron beam, thus converting to an $e^-e^-$ machine. This freedom allows us to study LNV processes such as $e^{-} e^{-} \to W^{-} W^{-} \to 4 j$, as discussed in Section~\ref{sec:2.5}. At the partonic level, 
there are no SM backgrounds for this LNV process. However, in realistic situations, different other processes can mimic the same final state as the signal. For instance, we can have $4j+\slashed{E}$, $4j+\slashed{E}+$ leptons or $4j+$ leptons, where the additional leptons are lost in the beam pipe. The SM processes which contribute to the background are listed in Ref.~\cite{Greub:1996ct}. We revisit these processes here and implicitly assume that we cannot reconstruct the charge of the gauge bosons from where the jets originated. The dominant background channels under consideration are : (P1) $e^-e^-\to W^- W^- \nu_{\ell} \nu_{\ell}$, (P2) $e^-e^-\to W^- Z \nu_{\ell} e^-$, (P3) $e^-e^-\to Z Z e^- e^-$ and (P4) $e^-e^-\to W^- W^+ e^- e^-$. All the gauge bosons decay hadronically so that we have four jets at the parton level. We shower the sample events and perform the detector simulation by demanding a minimum $\slashed{E}_T$ of $20$ GeV, no leptons and photons which pass the trigger cuts and at least $4$ tagged jets satisfying the trigger cuts given in Eq.~\eqref{jet_TC}. The background cross sections before and after the selection cuts are shown in Table~\ref{tab:ee24j-bkg}. For $\sqrt s=500$ GeV and $|V_{eN}|=0.04$, we can achieve $>3\sigma$ significance for the full range of the heavy neutrino mass considered here, as shown in Figure~\ref{fig:ee24j-sig}. However, given the smallness of the cross sections, the significance is somewhat lower at $\sqrt{s}=350$ GeV. We comment here that a more involved search strategy using optimized cuts or a multivariate analysis might improve this significance. Also note that, in our analysis, we have used unpolarized electron beams. However, with the experimental facility at the ILC, it will be possible to use polarized beams which will lead to an increase in the cross sections by a factor of four.

\begin{table}[!h]
\centering	
\begin{tabular}{|c|c|c|c|c|c|c|c|c|c|}
\hline 
$\sqrt{s}$  & \multicolumn{2}{|c|}{$\sigma_{P1}$ (fb)} & \multicolumn{2}{|c|}{$\sigma_{P2}$ (fb)} & \multicolumn{2}{|c|}{$\sigma_{P3}$ (fb)} & 
\multicolumn{2}{|c|}{$\sigma_{P4}$ (fb)} & $\mathcal{N}_B$\\ \cline{2-9}
(GeV) & TC & SC & TC & SC & TC & SC & TC & SC & \\
\hline 
350 & 0.067 & 0.004 & 0.048 & 0.001 & $\sim 0$ & $\sim 0$ & 0.132 & 0.005 & 5\\
\hline
500 & 0.453 & 0.015 & 0.399 & 0.004 & 0.006 & $\sim0$ & 0.718 & 0.025 & 22\\
\hline 
\end{tabular} 
\caption{Background cross sections for the signal process $e^-e^-\to W^-W^-\to 4j$ after TC and SC for the four processes mentioned 
in the text for $\sqrt{s}$ = 350 GeV and 500 GeV. The number of background events $\mathcal{N}_B$ is computed at an integrated luminosity of $500$ fb$^{-1}$.}
\label{tab:ee24j-bkg}
\end{table}

For completeness, we also show the partonic cross sections for all possible decay modes of the $W$-pair in Table~\ref{cross2tab}. Due to larger branching ratios for the hadronic decay of $W$, the $e^{-}e^{-} \to 4j$ has the largest cross section. Note that for the numbers in Table~\ref{cross2tab}, we have assumed the two $W$s to be on-shell. The numbers for the second and fourth column will not change much if any of the $W$s go off-shell; however, the numbers in the third column are valid only for on-shell $W$s.

\begin{table}[!h]
\centering	
\begin{tabular}{|c|c|c|c|}
\hline 
$M_N$ (GeV) & $4j$ (fb) & $2\ell^{-} +\slashed{E}$ (fb)  &  $\ell^{-}+2j+\slashed{E}$ (fb)   \\ 
\hline
250 & 0.075   &  0.009  & 0.055  \\
\hline
400 &  0.084  & 0.010  & 0.059   \\
\hline
\end{tabular}
\caption{Parton level signal cross sections of different channels arising from $e^-e^-\to W^-W^-$ after trigger and selection cuts for illustrative cases of the heavy neutrino masses $M_N=250$ and $ 400$ GeV and with mixing parameter $|V_{eN}|=0.04$ at $\sqrt s=500$ GeV.  }
\label{cross2tab}
\end{table}

\section{Multivariate Analysis}\label{sec:4} 

\begin{figure}[!h]
\centering
\includegraphics[scale=0.75,trim=0cm 0cm 0cm 0cm,clip=true]{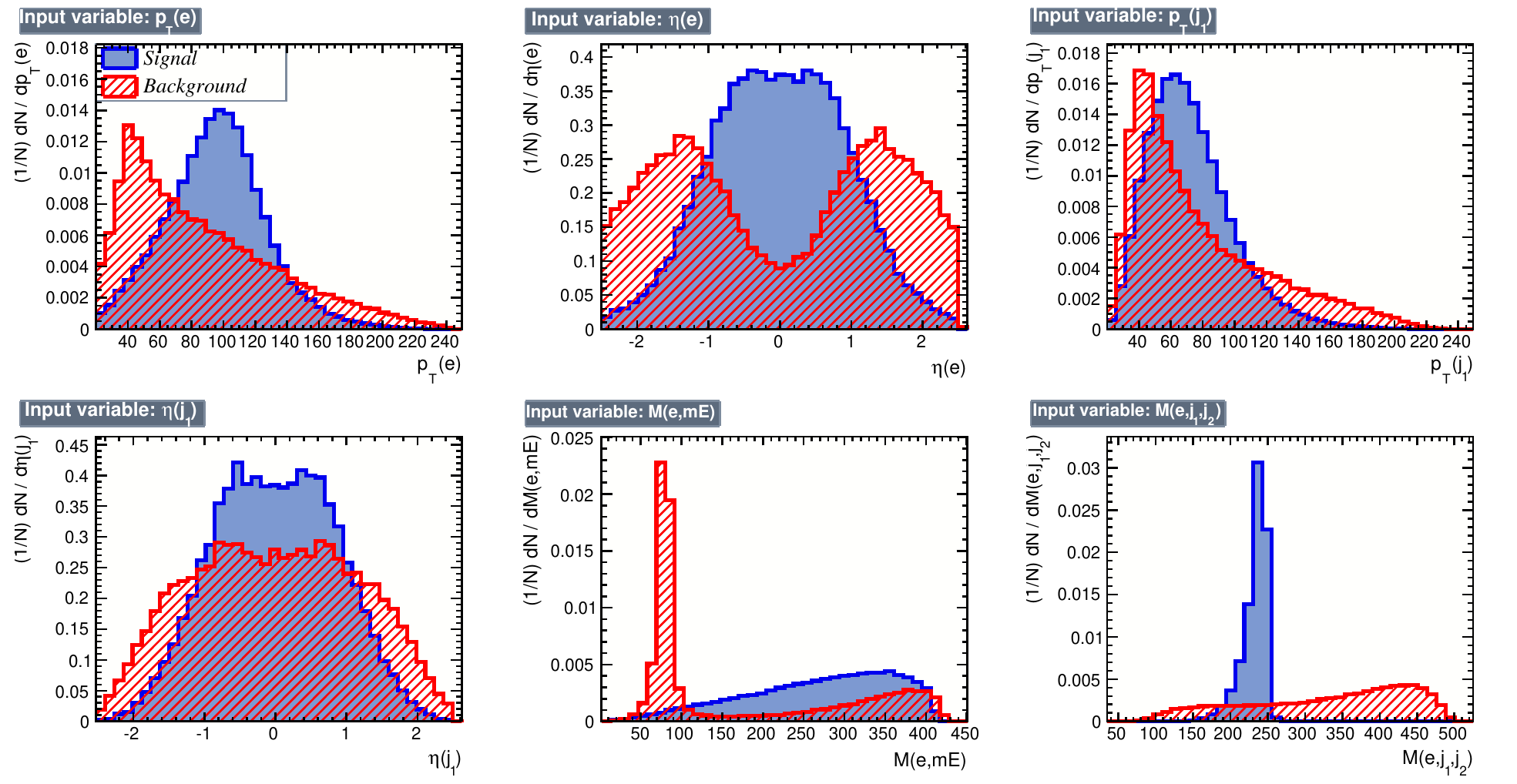}
\includegraphics[scale=0.75,trim=0cm 5.4cm 0cm 0cm,clip=true]{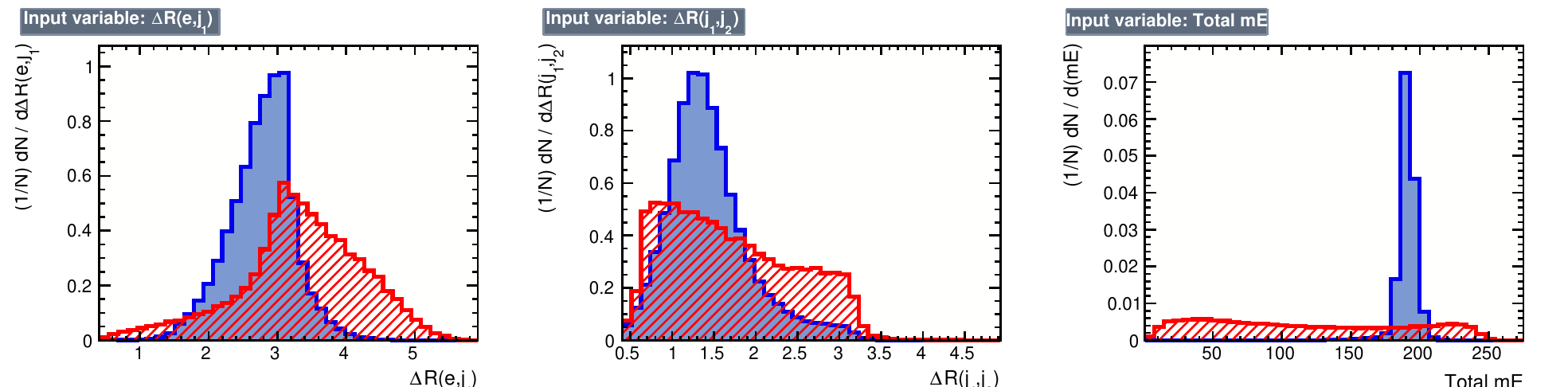}
\caption{Signal (blue) and background (red) distributions of the input variables used for MVA. All the distributions are drawn for $M_N=250$ GeV, $|V_{eN}|=0.04$ and $\sqrt{s}=500$ GeV.}
\label{fig:varBDT}
\end{figure}

In this section, we present a multivariate analysis for better signal-to-background discrimination, which leads to a better significance. For MVA, we use the Boosted Decision Tree (BDT) algorithm in the~\textsc{TMVA}~\cite{Hocker:2007ht} framework. 
In the cut-based analysis for the channel $e^+e^-\to N\nu_e\to e+2j+\slashed{E}$ presented in Section~\ref{subsec:ee2Nv}, we have shown that two simple invariant mass cuts [cf.~Eq.~(\ref{eq:SCCBA})] are sufficient to isolate the signal from the background with $\gtrsim 5\sigma$ significance, particularly for larger mixing ($|V_{eN}| \gtrsim 0.02$) and in smaller mass range ($\lesssim 350$ GeV). To probe smaller mixing ($|V_{eN}| \lesssim 0.02$), we need to find  an optimized set of cuts for the CBA in order to achieve 5$\sigma$ significance. For this purpose, the MVA technique is a useful tool to obtain the best sensitivity for a given set of parameters. For the MVA in the present case, we use nine variables, as listed in Table~\ref{tab:MVAvar} with their relative importance in the MVA output response. These variables are chosen by comparing the signal trained for $M_N=250$ GeV with $|V_{eN}|=0.04$ at $\sqrt{s}=500$ GeV with the background distributions. Each of these variables has some reasonable amount of discriminating power, as shown in Figure~\ref{fig:varBDT}, where we have plotted the normalized signal and background distributions. It is important to mention here that this set of nine variables used here might not be the optimal one and there is always a scope to improve the analysis with cleverer choices of variables. In our analysis, we have used these simple kinematic variables which are less correlated and have sufficiently good discriminating power. We use the MVA input-variables set optimized for the parameters $M_N=250$ GeV, $|V_{eN}|=0.04$ and $\sqrt{s}=500$ GeV for all parameter choices. In a more dedicated MVA, one can use different set of variables for different parameter points to further improve the analysis. 
\begin{table}[!h]
\centering
\begin{tabular}{|c|c|c|c|c|c|}
\hline 
Variable & Importance & Variable & Importance & Variable & Importance \\ 
\hline 
$p_T(e)$    & $9.123 \times 10^{-2}$ & $|\eta(j_1)|$ & $6.704 \times 10^{-2}$ & $\Delta R(e,j_1)$ & $5.662 \times 10^{-2}$ \\ 
\hline 
$|\eta(e)|$ & $7.824 \times 10^{-2}$ & $M(e,\slashed{E})$ & $1.527 \times 10^{-1}$ & $\Delta R(j_1,j_2)$ & $6.180 \times 10^{-2}$ \\ 
\hline 
$p_T(j_1)$  & $7.766 \times 10^{-2}$ & $M(e,j_1,j_2)$ & $2.522 \times 10^{-1}$ & Total $\slashed{E}$ & $1.625 \times 10^{-1}$ \\ 
\hline 
\end{tabular} 
\caption{Input variables used for MVA with $M_{N}=250$ GeV, $|V_{eN}|=0.04$ and $\sqrt{s}=500$ GeV and their relative importance. This might vary for other sets of parameters.}
\label{tab:MVAvar}
\end{table}
\begin{figure}[!h]
\centering
\includegraphics[scale=0.50]{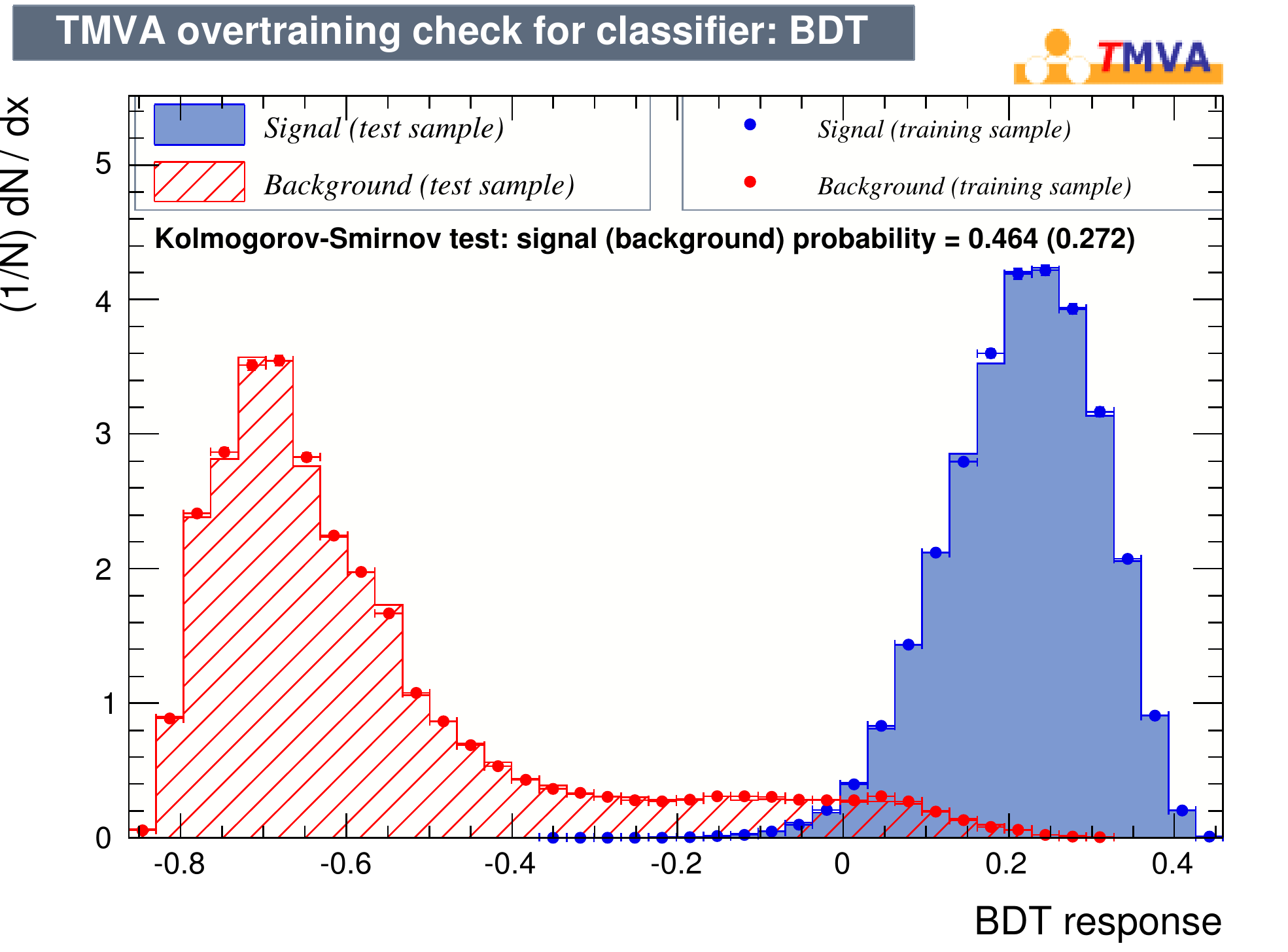}
\caption{BDT overtraining check for the parameter set $M_{N}=250$ GeV, $V_{eN}=0.04$ and $\sqrt{s}=500$.}
\label{fig:OT}
\end{figure}

For MVA, we must always be very careful about overtraining the signal/background, since it can happen without the proper choices of the algorithm specific tuning parameters.  Using the Kolmogorov-Smirnov (KS) test one can check whether a test sample is overtrained or not. Generally, the test sample is not overtrained if the KS probability lies within the range 0.1 to 0.9. For most cases, a critical KS probability value greater than 0.01~\cite{KS-test} ensures that the samples are not overtrained. In Figure~\ref{fig:OT}, we have shown the KS probability values for the signal and the background of the BDT response and this confirms that both signal and background samples are not overtrained. We have ensured that we do not encounter overtraining for the full parameter range. We have used two statistically independent samples for MVA, one for training and the other for testing. As can be seen from Figure~\ref{fig:OT}, the signal and background samples are well separated in this BDT output, and hence, using an appropriated BDT cut, we can significantly increase the signal significance.

In Table~\ref{tab:MVA1}, we compute the significance for different $M_N$ and fixed $|V_{eN}|=0.04$ at $\sqrt{s}=500$ GeV with $\mathcal{L}=$ 100 fb$^{-1}$. In Table~\ref{tab:MVA2}, the significance is computed for different mixing values for a fixed $M_N=250$ GeV at $\sqrt{s}=500$ GeV with $\mathcal{L}=$ 100 fb$^{-1}$. From these two tables, it is evident that MVA is more powerful to probe smaller mixing and larger masses where the performance of the CBA is comparatively weaker.

\begin{table}[!ht]
\centering
\begin{tabular}{|c|c|c|c|c|c|c|}
\hline 
$M_{N}$ & $\sigma^{\rm w/o~cut}$ & $\sigma^{\rm TC}$ & BDT cut & $\mathcal{N}_{S}$ & $\mathcal{N}_{B}$ & $\mathcal{N}_S$   \\ 
(GeV) & (fb) & (fb) & value   &  &    & $\overline{\sqrt{\mathcal{N_S}+\mathcal{N_B}}}$\\
\hline 
100 & 39.28 & 9.917 & 0.162 &  926 &  301 & 26.4 \\ 
\hline 
150 & 38.40 & 22.61 & 0.162 & 1778 &  946 & 34.1 \\ 
\hline 
200 & 34.06 & 22.73 & 0.159 & 1754 &  786 & 34.8 \\ 
\hline
250 & 30.22 & 21.94 & 0.156 & 1764 &  651 & 35.9 \\ 
\hline 
300 & 25.79 & 19.87 & 0.136 & 1659 &  475 & 35.9 \\ 
\hline 
350 & 20.42 & 16.33 & 0.167 & 1315 &  431 & 31.5 \\ 
\hline 
400 & 13.95 & 11.43 & 0.232 &  876 &  294 & 25.6 \\ 
\hline 
450 & 6.605 & 5.000 & 0.218 &  297 &  194 & 13.4 \\ 
\hline 
475 & 2.815 & 2.361 & 0.280 &  119 &  357 &  5.5 \\ 
\hline \hline 
SM & 1022.5 & 638.3 & - & - & - & - \\ 
\hline 
\end{tabular} 
\caption{Partonic cross sections ($\sigma^{\rm w/o~cut}$) and cross sections after TC ($\sigma^{\rm TC}$) as defined in Section~\ref{subsec:ee2Nv} for the CBA with detector effects included for the process $e^+e^-\to N\nu_e\to e+2j+\slashed{E}$ at $\sqrt{s}=500$ GeV with mixing $|V_{eN}|=0.04$. Numbers of 
signal ($\mathcal{N}_S$) and background ($\mathcal{N}_B$) events for different $M_N$ after passing the BDT cut are shown for luminosity $\mathcal{L}=$ 100 fb$^{-1}$.}
\label{tab:MVA1}
\end{table}
\begin{table}[!h]
\centering
\begin{tabular}{|c|c|c|c|c|c|c|}
\hline  
$|V_{eN}|$ & $\sigma^{\rm w/o~cut}$ & $\sigma^{\rm TC}$ & BDT cut & $\mathcal{N}_{S}$ & $\mathcal{N}_{B}$ & $\mathcal{N}_S$   \\ 
 & (fb) & (fb) & value  &  &  & $\overline{\sqrt{\mathcal{N_S}+\mathcal{N_B}}}$ \\
\hline 
0.030 & 17.00 & 11.90 & 0.158 & 938 & 580 & 24.1 \\ 
\hline 
0.025 & 11.81 &  8.27 & 0.201 & 582 & 360 & 19.0 \\ 
\hline 
0.020 &  7.55 &  5.29 & 0.227 & 325 & 248 & 13.6 \\ 
\hline 
0.015 &  4.25 &  2.98 & 0.220 & 187 & 303 & 8.4 \\ 
\hline 
0.010 &  1.89 &  1.32 & 0.336 &  47 &  60 & 4.6 \\ 
\hline
0.0075 & 1.06 &  0.74 & 0.309 &  28 &  71 & 2.8 \\ 
\hline 
0.0050 & 0.47 &  0.33 & 0.359 &   8 &  30 & 1.3 \\
\hline 
\end{tabular} 
\caption{Same as Table~\ref{tab:MVA1} for different mixing $|V_{eN}|$ and fixed $M_N=250$ GeV.}
\label{tab:MVA2}
\end{table}
We also analyze the two subdominant channels $e^+e^-\to N\nu_e\to Z\nu_e\nu_e\to 2j+\slashed{E}$ and $e^+e^-\to N\nu_e\to H\nu_e\nu_e\to 2b+\slashed{E}$ for a few illustrative masses
using a simplified MVA. For these two channels we have used eight variables, viz. $p_T(j_1)$, $|\eta(j_1)|$, $\slashed{E}_T$, $|\eta(\slashed{E}_T)|$,
$M(j_1,j_2)$, $M(\slashed{E},j_1)$, $\Delta\eta(\slashed{E},j_1)$ and scalar sum $H_T$ of the $p_T$'s of all visible particles,  to distinguish signal from background.
Before the training and testing of the samples, they were made to pass the trigger cuts given in Section~\ref{subsec:ee2Nv}.  In Table~\ref{tab:NZ} for the channel 
$e^+e^-\to N\nu_e\to Z\nu_e\nu_e\to 2j+\slashed{E}$, we show estimates for significance for three illustrative mass values, viz. $M_N=100,200$ and $300$ GeV for a fixed mixing 
$|V_{eN}|=0.04$ at $\sqrt{s}=350$ GeV. Similarly, in Table~\ref{tab:NH} we show analogous results for the channel $e^+e^-\to N\nu_e\to H\nu_e\nu_e\to 2b+\slashed{E}$ with mass values 
$M_N=150,200$ and $300$ GeV for a fixed mixing $|V_{eN}|=0.04$ at $\sqrt{s}=350$ GeV. It is clear that using MVA, one can even probe these two subdominant channels with a good significance. 
It is important to note here that the significance for the process $e^+e^-\to N\nu_e\to H\nu_e\nu_e\to 2b+\slashed{E}$ is subdued because of the smaller branching ratio of $N\to H \nu$ than 
that of $N\to Z \nu$ for the benchmark points under consideration (see Figure~\ref{fig:BR}). Also  we consider a single $b$-jet tagging in order to identify the Higgs, and this further reduces the signal due to the $b$-tagging efficiency, which could vary from about 40\% for low $p_T$ $b$-jets to about 70\% for high $p_T(b)$. By demanding two $b$-tags, the significance will decrease even further. Once the experiments start running and the detector performances are better understood, one can try to look for cleverer variables to improve this 
significance.

\begin{table}[!h]
\centering
\begin{tabular}{|c|c|c|c|c|c|c|}
\hline 
$M_{N}$ & $\sigma^{\rm w/o~cut}$ & $\sigma^{\rm TC}$ & BDT cut & $\mathcal{N}_{S}$ & $\mathcal{N}_{B}$ & $\mathcal{N}_S$   \\ 
(GeV) & (fb) & (fb) & value   &  &    & $\overline{\sqrt{\mathcal{N_S}+\mathcal{N_B}}}$\\
\hline 
100 &  8.27 & 5.89 & 0.026 & 480 & 8555 & 5.1 \\
\hline
200 & 13.61 & 10.22 & 0.017 & 724 & 6938 & 8.3 \\
\hline
300 & 4.39  & 3.738 & 0.082 & 275 & 5464 & 3.6 \\
\hline \hline
SM & 290.0 & 218.9 & - &- &-  &-  \\
\hline
\end{tabular}
\caption{Same as Table~\ref{tab:MVA1} for the process $e^+e^-\to N\nu_e\to Z\nu_e\nu_e\to 2j+\slashed{E}$ at $\sqrt{s}=350$ GeV. }
\label{tab:NZ}
\end{table}
\begin{table}[!h]
\centering
\begin{tabular}{|c|c|c|c|c|c|c|}
\hline
$M_{N}$ & $\sigma^{\rm w/o~cut}$ & $\sigma^{\rm TC}$ & BDT cut &
$\mathcal{N}_{S}$ & $\mathcal{N}_{B}$ & $\mathcal{N}_S$   \\
(GeV) & (fb) & (fb) & value   &  &    &
$\overline{\sqrt{\mathcal{N_S}+\mathcal{N_B}}}$\\
\hline
150 &  2.24 & 0.795 & 0.243 & 36 & 260 & 2.1 \\
\hline
200 &  5.49 & 1.952 & -0.058 & 149 & 1580 & 3.6 \\
\hline
300 &  2.79  & 1.045 & -0.005 & 82 & 1621 & 2.0 \\
\hline \hline
SM & 290.0 & 44.45 & - & -& - &-  \\
\hline
\end{tabular}
\caption{Same as Table~\ref{tab:MVA1} for the process $e^+e^-\to N\nu_e\to H\nu_e\nu_e\to 2b+\slashed{E}$ (only one
$b$-jet is tagged among two $b$'s) for c.m. energy $\sqrt{s}=350$ GeV. The corresponding trigger cuts are defined in Section~\ref{subsec:ee2Nv}.}
\label{tab:NH}
\end{table}
\section{Conclusion} \label{sec:conclu} 
The seesaw mechanism provides a simple understanding of the observed smallness of the neutrino masses. Heavy Majorana neutrinos are an essential ingredient of the simplest seesaw scenario, and therefore, it is important to explore their experimental signatures in current and future experiments. In this work, we have studied the prospects of heavy neutrino searches at future lepton colliders. Due to their relatively cleaner environment, as compared to hadron colliders, lepton colliders provide us with a unique opportunity to probe the heavy neutrinos with higher sensitivity, which might shed light on the origin of neutrino mass and the associated new physics beyond the SM. In view of the increasingly hopeful case for an International Linear Collider in near future, we have mainly focused on various heavy neutrino production mechanisms at an $e^+e^-$ collider and have performed a detailed detector-level simulation for the most promising channels. In particular, 
the production process $e^+e^-\to N\nu$ leading to the final state $e+2j+\slashed{E}$ has the largest cross section, and hence, we analyze this possibility in detail, taking into account realistic detector efficiency and SM background estimates.  We have followed two different approaches to separate the signal from background. First, we present a conventional cut-based analysis using the distinct kinematic features of the signal over background. We show that in this approach, a light-heavy neutrino mixing parameter $|V_{eN}|^2 \sim 10^{-4}$ can be probed at 95\% C.L. for a heavy neutrino mass up to $400$ GeV at $\sqrt s=500$ GeV ILC with luminosity of  $100~\rm{fb}^{-1}$. Then we use a multivariate analysis with a boosted decision tree algorithm to make a better discrimination of the signal from background, which enables us to probe even smaller mixing values up to $|V_{eN}|^2 \sim 10^{-5}$ or so for a larger range of heavy neutrino masses.  

The dominant production mode $e^+e^-\to N\nu$ does not probe the Majorana nature of heavy neutrinos, which is a crucial aspect of the seesaw mechanism. To overcome this shortcoming at lepton colliders, we have studied a new production channel: $e^+e^-\to N\ell^\pm W^\mp$, leading to a lepton number violating same-sign dilepton plus four jet final state. Although the production cross section for this process is smaller than that of $e^+e^-\to N\nu$, the clean environment of $e^+e^-$ colliders makes this process almost background free. Thus, any positive detection in this channel would be strong evidence for lepton number violation, with profound implications for our understanding of the neutrino mass mechanism. We find that a detection of 10 signal events in this channel for $|V_{eN}|=0.04$ would require an integrated luminosity of about 700 fb$^{-1}$ for $\sqrt s=500$ GeV, which is certainly within reach of next generation lepton colliders.  

Finally, we also study the lepton number violating process $e^{-} e^{-} \to W^{-} W^{-}$, mediated by a $t$-channel heavy neutrino exchange, followed by the hadronic decay of the $W$s. This channel is again almost background-free. Hence, a positive observation of this signal will be an unambiguous probe of the Majorana nature of the heavy neutrinos. We show that in this channel, we can achieve $>3\sigma$ significance for the full range of the heavy neutrino mass considered here at a $\sqrt{s}=500$ GeV ILC operating in the $e^-e^-$ mode with a luminosity of $\mathcal{L}=500 \, \rm{fb}^{-1}$, provided the mixing is relatively large.  With polarized beams, the signal cross section can be enhanced by a factor of four, thus leading to a better sensitivity.

Although our numerical simulations have been done specifically with the ILC in mind, our analysis can be easily extended to other next generation lepton colliders, such as the CLIC, FCC-ee, CEPC or even muon-muon, electron-photon and photon-photon colliders, once the design parameters and detector specifics become clearer and readily available for these machines.  

\section*{Acknowledgments} 

S.B. and T.M. thank Satyaki Bhattacharya and Shilpi Jain for their help regarding the multivariate analysis. P.S.B.D. thanks the local hospitality provided at TUM, Munich where part of this work was done. The work of S.B. and T.M. was partially supported by funding available from the Department of Atomic Energy, Government of India, for the Regional Centre for Accelerator-based Particle Physics (RECAPP), Harish-Chandra Research Institute. 
The work of P.S.B.D. is supported by the Lancaster-Manchester-Sheffield Consortium
for Fundamental Physics under STFC Grant No. ST/L000520/1. The work of A.I. was done in the context of the ERC Advanced grant project ``FLAVOUR'' (Grant No. 267104) and was partially supported by the DFG cluster of excellence ``Origin and Structure of the Universe''.  M.M. acknowledges partial support of the ITN INVISIBLES (Marie Curie Actions, Grant No. PITN-GA-2011-289442).

\end{document}